\documentclass[]{emulateapj}

\usepackage{longtable}
\usepackage{natbib}
\usepackage{graphicx}
\begin{document}

\title{A simple connection between the near- and mid-infrared emission of galaxies and their star-formation rates}

\author{\medskip 
Erin Mentuch\altaffilmark{1},
Roberto G. Abraham\altaffilmark{1},
Stefano Zibetti\altaffilmark{2}}

\altaffiltext{1}{Department of Astronomy \& Astrophysics, University of Toronto, 50 St. George Street, Toronto, ON, M5S~3H4, Canada}
\altaffiltext{2}{Max-Planck-Institut fur Astronomie, Konigstuhl 17, D-69117 Heidelberg, Germany}

\begin{abstract}

We have measured the near-infrared colors and the fluxes of individual pixels in 68 galaxies common to the \textit{Spitzer} Infrared Nearby Galaxies Survey and the Large Galaxy Atlas Survey. Each galaxy was separated into regions of increasingly red near-infrared colors. In the absence of dust extinction and other non-stellar emission, stellar populations are shown to have relatively constant NIR colors, independent of age.  In regions of high star formation, the average intensity of pixels in red-excess regions (at 1.25\,\micron, 3.6\,\micron, 4.5\,\micron, 5.6\,\micron, 8.0\,\micron~and 24\,\micron) scales linearly with the intrinsic intensity of H$\alpha$ emission, and thus with the star-formation rate within the pixel. This suggests that most NIR-excess regions are not red because their light is being depleted by absorption. Instead, they are red because additional infrared light is being contributed by a process linked to star-formation. This is surprising because the shorter wavelength bands in our study (1.25\,\micron--5.6\,\micron) do not probe emission from cold (10--20~K) and warm (50--100~K) dust  associated with star-formation in molecular clouds. However, emission from {\em hot} dust (700--1000~K) and/or Polycyclic Aromatic Hydrocarbon molecules can explain the additional emission seen at the shorter wavelengths in our study. The contribution from hot dust and/or PAH emission at 2\,\micron--5\,\micron~and PAH emission at  5.6\,\micron~and 8.0\,\micron~scales linearly with warm dust emission at 24\,\micron~and the intrinsic H$\alpha$ emission. Since both are tied to the star-formation rate, our analysis shows that the NIR excess continuum emission and PAH emission at $\sim1-8\,$\micron~can be added to spectral energy distribution models in a very straight-forward way, by simply adding an additional component to the models that scales linearly with star-formation rate.

\end{abstract}

\section{Introduction}

The signatures of star formation are expressed across a galaxy's spectral energy distribution (SED). Ultraviolet and optical light trace the recently formed stellar populations, infrared light maps out luminous dust grains that re-emit this same starlight, and radio emission traces magnetic fields produced by short-lived, massive stars. How well these tracers can be used to determine a galaxy's star-formation rate depends on the galaxy's luminosity, dust content, nuclear activity, and metallicity. Adding to the complexity, the timescale of each tracer depends on the emission mechanism, hydrogen is ionized by only the most massive stars on timescales of $10^7$ years, while dust emission can be heated by lower mass, longer-lived stars on the order of several $10^8$ years. At a fundamental level these tracers
are highly interdependent, because energy must be conserved. For example, the energy in the UV radiation absorbed by dust must be balanced out by the re-emission of this energy at other wavelengths. This suggests that the best estimator for star-formation rate will be based on a combination of tracers which together capture the full energetics of the system.

Recent work by \citet{cal07} and \citet{ken09} shows that a good estimator for the star-formation rate emerges from the linear combination of a galaxy's near-UV or visible-wavelength emission (either UV continuum or line fluxes, attenuated by dust) and its emission in the infrared (e.g. 8\,\micron, 24\,\micron, or total IR\,$\lambda$ 8--1000 and also radio). The beauty of this approach is that the data itself can correct for the attenuation and emission of dust. 

At near-infrared (NIR) (0.9-5\,\micron) and mid-infrared (MIR) wavelengths (5-30\,\micron), emission due to heating of very small grains (VSG) and polycyclic aromatic hydrocarbons (PAH) also contributes to the spectral energy distribution. UV photons excite these small molecules which emit through various vibrational/stretching/rotational modes and thus this emission should trace regions of star formation. PAH emission around 8\,\micron~shows spatial correlation to star formation within \ion{H}{2} regions \citep{cal07}, and can be used as a SFR indicator, particularly if used in combination with an optical SFR indicator \citep{ken09}. Observations of nearby star forming galaxies suggest that the spectrum of PAH line emission is expressed through constant line ratios. For example, \citet{lu03} found constant ratios of stellar subtracted flux densities at 4, 6.2, 7.7 and 11.3. PAH and/or dust emission at shorter wavelengths has received less attention, partly because of the stronger contribution from stellar continuum emission at these wavelengths.

In addition to PAH line emission, NIR continuum emission in excess of NIR stellar emission can be significant at wavelengths of 2-5\,\micron. High redshift star forming galaxies, where SFRs are significantly higher than nearby, emit significant NIR excesses at this wavelength \citep{men09,mag08} and its emission correlates with optical SFR indicators \citep{men09}. Nearby galaxies, studied with the \textit{Infrared Space Observatory} also show NIR excess continuum emission \citep{lu03} which scales linearly with the PAH spectrum and shows trends of increasing with increasing star formation indicators. This NIR dust emission component is seen not only in extragalactic emission, but has been found locally in galactic cirrus emission \citep{fla06}, reflection nebulae \citep{sel83,sel96}, planetary nebulae \citep{der06} and massive star-forming regions \citep{mae05,mae06,lon07}. In the latter two cases, the NIR dust emission is linked to circumstellar disk emission \citep{woo08,Touhami:2010p4410,Acke:2004p4409} where high luminosity OB stars heat dust to its sublimation temperature (for example a $10^5$\,L$_\odot$star will heat dust to 1000\,K at r = 7.3\,AU, while a $10^2$\,L$_\odot$star will heat dust to 1000\,K at r = 0.23\,AU and a L$_\odot$ star will heat dust up to 1000\,K out to a radius of 0.023\,AU.) 

The source of extragalactic NIR dust emission, and its contribution to a galaxy's integrated light remains uncertain. The excess seen in higher star forming galaxies at high redshift is possibly a combination of both NIR dust emission related to star formation regions (which includes reflection nebulae and circumstellar dust emission) and NIR emission due to cirrus emission. In \citet{men09}, we show that there is a trend of increasing NIR continuum emission with star formation and show that the magnitude of NIR excess emission can be well matched if the majority of the excess NIR emission originates in hot (700~K--1000~K) thermal emission, possibly from dust in circumstellar disks heated to their sublimation temperature. Other sources of NIR emission such as dust heated by active galactic nuclei, planetary nebula and reflection nebula can also be important, although their contribution to the integrated light of a galaxy is expected to be less significant \citep{men09}.

A central purpose of the present paper is to test this hypothesis, using the images of nearby galaxies in which we are able to resolve the components of the galaxy at greater resolution, giving us the opportunity to determine if the NIR excess emission is associated with emission from young stellar populations, galactic cirrus or active galactic nuclei. Additional narrow-band H$\alpha$ imaging, which traces the young, massive stellar populations,  complements the analysis. We perform multiwavelength pixel-by-pixel analysis on a sample of 68 nearby galaxies from the \textit{Spitzer} Infrared Nearby Galaxies Survey (SINGS; \citealt{ken03}) with complementary near-infrared data from the Large Galaxy Atlas (LGA; \citealt{jar03}). These data resolve galaxies on physical scales between 8\,pc (for the closet galaxy NGC\,6822 at a distance of 0.6\,Mpc) and $\sim400$\,pc (for the most distant galaxies at $\sim$30\,Mpc). 

Near-infrared colors of evolved stellar populations are not highly sensitive to a population's age, although contributions from post-main-sequence stars can contribute at ages of $\sim$1-3\,Gyr. The NIR traces the Rayleigh-Jeans tail of stellar emission and as a result the colors remain nearly constant as stars of progressively lower masses leave the main sequence. However, observations have shown that in addition to emission in the NIR, extinction due to dust leads to reddened colors as is shown in \citet{ind05} in our own galaxy and in nearby starbursts \citep{Calzetti:1994p1499}. Integrated colors consist of light from all the stellar populations and dust in a galaxy, making it difficult to discern whether red NIR colors are associated with non-stellar emission or are merely due to dust extinction. Our study provides synergy between investigations of integrated emission for whole galaxies, and studies of resolved stellar populations by grouping pixels with common near-infrared colors together so that the common properties of these regions are revealed. 

We motivate the need for pixel-by-pixel color studies in the following section. The rest of the paper is outlined as follows. In \S\,\ref{s:data}, we briefly summarize our sample selection from the SINGS and LGA surveys. Our pixel-by-pixel color selection and image processing is described in \S\,\ref{s:pp}. Results from the average NIR colors and intensities at H$\alpha$, 1.25\,\micron, 3.6\,\micron, 4.5\,\micron, 5.6\,\micron, 8.0\,\micron~and 24\,\micron~measured in each color selected region are presented in \S\,\ref{s:results}. Implications for studying integrated galaxies and applying our results to SED models are discussed in \S\,\ref{s:disc}.

\section{Pixel-by-pixel NIR colors of resolved galaxies}\label{s:mot}

An analysis of galaxy colors at the level of individual pixels is
substantially more complicated than the usual approach based on
integrated photometry using large
apertures. It is important to establish from the outset
whether such a complicated approach is really required in order to understand the underlying physics. 
Before embarking on a description of our methodology, we will first attempt to
justify the underlying need for it.

At near-ultraviolet and visible wavelengths a galaxy's spectral energy distribution is the sum of the light from the stellar populations 
that make up the galaxy,
modulated by dust absorption, and augmented by additional sources of emission (such as from nebular
lines). The different
ages of the stellar populations impact the SED and the resulting colors, as young high-mass stars peak in their emission at shorter wavelengths and have blue visible colors, while older low-mass stars peak in their emission at red wavelengths and have redder visible colors. Slightly longer near-infrared wavelengths (around $\sim$1--2\,\micron) are ideal for probing the bulk of a galaxy's stellar mass, because starlight still dominates the overall SED
and most of the light at these
wavelengths comes from late evolutionary stages of low-mass stars which dominate the stellar mass budget.
The colors of stellar populations in the near-infrared are relatively constant for most stellar spectral types and most evolutionary phases, with few notable exceptions like the TP-AGB stars (e.g. \citealt{mar05}).
Slightly longer wavelengths, say between 2\,\micron~and 5\,\micron, are special. This wavelength range corresponds to a minimum in a galaxy's SED, because the Rayleigh-Jeans tail of starlight is rapidly declining but thermal emission from warm dust is not yet
important. Additional components of a galaxy that happen to emit at these wavelengths, such as hot
dust, may be detectable at these wavelengths even if they do not contribute a large fraction of a galaxy's bolometric output,
simply because of contrast against the weak underlying galactic SED.
At wavelengths longer than 5\,\micron~starlight starts to play a negligible role, and the physics of dust emission takes over.
For example in the \textit{Spitzer} 5.6\,\micron~and  8.0\,\micron~bands, molecular emission lines from Polycyclic Aromatic Hydrocarbons (PAHs) are important, while at 24\,\micron~most of the light comes from the thermal emission of warm dust. 

\begin{figure*}[tbp]
\begin{center}
\includegraphics[width=6in]{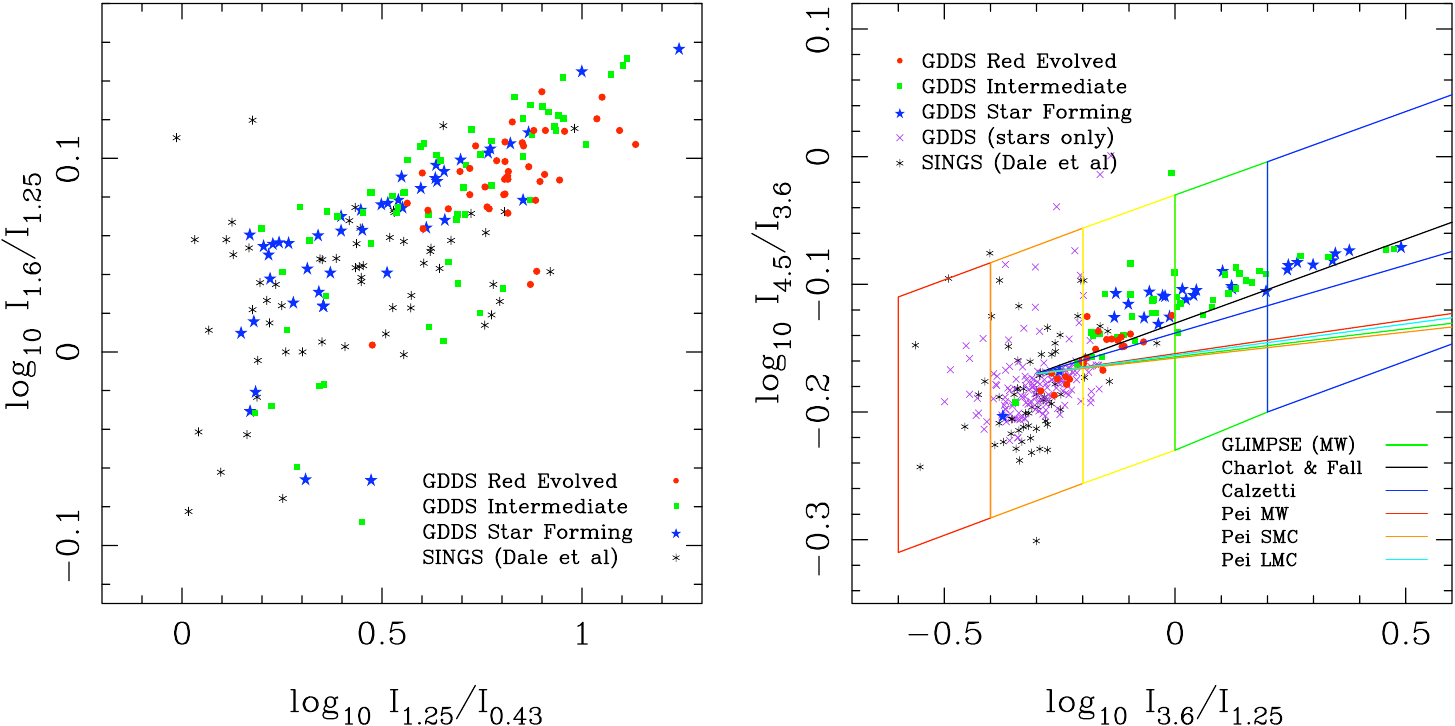}
\caption[NIR and optical/NIR Color-color diagrams for integrated galaxy photometry from SINGS and GDDS surveys]{Color-color plots in the optical/NIR (left) and the NIR (right) for the integrated SINGS photometry from \cite{dal07} and rest-frame colors for the GDDS high-redshift galaxy sample. Rest-frame colors are computed using best fit SEDs from \citet{men09}. The spread in I$_{1.25}/$I$_{0.43}$ color (left panel) results from variations in both age and dust attenuation and is well described by SPS models, while the spread in I$_{3.6}/$I$_{1.25}$ color (right panel) is largest for high SFR galaxies, but is not well described by SPS models (purple points) although SPS models augmented with NIR emission (colored points, keyed to legend according to galaxy spectral type) provide a better model. In the right panel, extinction vectors are plotted from the color locus of (log$_{10}$ I$_{3.6}$/I$_{1.25}$, log$_{10}$ I$_{4.5}$/I$_{3.6}$ = -0.30, -0.17) for various dust extinction laws. The colored boxes indicate regions of increasing NIR color, defined in \S\,\ref{s:pp} and used as color selection in the quantitative analysis presented in this analysis.}
\label{fig:intcolors}
\end{center}
\end{figure*}

Some of these ideas are illustrated in Figure~\ref{fig:intcolors}.  The left-hand panel of the figure
shows a ratio of optical and near-infrared colors. Integrated photometry from \citet{dal07} for the SINGS sample (black asterisks) is plotted along with rest-frame colors from a high-redshift galaxy sample (data taken from the Gemini Deep
Deep Survey; \citealt{bob04}). For the high-redshift data, the colors were computed from the best fit SEDs from \citet{men09} and have been corrected for full cosmological dimming. Their symbols have been keyed to differing galaxy types as indicated by the spectral indicators given in \citet{bob04}. Star forming galaxies show spectral indicators of young stellar populations, while evolved galaxies are mainly comprised of older stellar populations. Intermediate galaxies show signatures of both evolved and young stellar populations. The axes can be converted to AB magnitude colors by multiplying by 2.5. 

The spread in the I$_{1.25}$/I$_{0.43}$ ratio (or, in magnitude space, $B-J$ color) is over 1 dex and is much larger than the spread in the NIR intensity ratios ($J-H$ color).  Star forming galaxies tend to have bluer $B-J$ color and evolved galaxies have redder $B-J$ color, but the intermediate galaxies are spread throughout (dust can play a major role in this spread). In the right-hand panel, near-infrared colors are plotted for both the SINGS and GDDS rest-frame sample. The rest-frame colors of the high redshift data are interpolated from SED models fit to the observed photometry. The purple crosses are from SED models that only incorporate emission from star light and attenuation by dust. The NIR colors of these models colors occupy a compact region in the NIR color-color plot. However, in \citet{men09}, we show that these models are not a good description of the near-infrared observed photometry and find that a SED stellar population augmented with a NIR emission model better represents the observations. Rest-frame colors derived from these SEDs are shown in the right panel (keyed to galaxy spectral types). These rest-frame colors of the high-redshift GDDS sample are consistent with the SINGS photometry for galaxies composed primarily of evolved populations, but for the star forming and intermediate GDDS galaxies, which have higher SFRs compared to the low redshift SINGS dataset, show a large spread toward increasing I$_{3.6}$/I$_{1.25}$ ratios (or bluer $J-L$ color) and  I$_{4.5}$/I$_{3.6}$ ratios. 

On the whole, Figure~\ref{fig:intcolors} makes three main points which motivate the analysis in this paper. (1) The purely
stellar component of the SED (derived from fitting of SPS models,
purple crosses in the right panel) shows a small ($\sim$0.2 dex) variation in
the NIR which is much smaller than at visible wavelengths, where the
variation is $\sim$1 dex.
(2) In spite of the previous point, 
real galaxies show much greater variation ($\sim 1$~dex) in near-infrared color than can be encompassed by stellar population models. (3) As a result
of the previous two points, the large variation in the NIR galaxy colors must be telling us quite a lot about
dust or some other source of non-stellar emission. Either the starlight is being reddened by dust absorption, or dust emission must be contributing to the colors
at surprisingly short wavelengths. 

Resolved galaxy colors can help distinguish between the latter two possibilities, because the properties of dust are known to vary from place-to-place in galaxies.
Figure~\ref{fig:intcolors} (right panel) shows that galaxy colors tend to scatter along the extinction vectors found in starburst regions, but {\em not} along the vectors  appropriate for regions outside of starburst regions. The extinction curve for lines of sight through our own galaxy \citep[obtained by the GLIMPSE survey,][]{ind05} is shown in green. Using the prescriptions presented in 
\cite{pei92}, we see that this extinction curve is consistent with the dust extinction law for the Milky Way (shown in red), the Small Magellanic Cloud (shown in orange) and the Large Magellanic Cloud (shown cyan). The extinction measured in starburst galaxies from \citet{Calzetti:1994p1499} is plotted in blue, and the effective dust extinction law from \citet{Charlot:2000p1497} is shown in black. In this figure we show all extinction vectors originating from the colors (log$_{10}$ I$_{3.6}$/I$_{1.25}$, log$_{10}$ I$_{4.5}$/I$_{3.6}$ = -0.30, -0.17) which we found to be the average colors of stellar emission in the NIR from the PEGASE.2 stellar SEDs\footnote{These colors are likely redder than that from pure stellar emission. Attenuation due to dust was included in these models. We later show (in \S\,\ref{s:results}) that these colors are consistent with the median pixel color of the majority of pixels which compose the SINGS galaxy sample.}.  The trend for NIR colors along this vector, predicted to be due to dust extinction in starburst regions, is supported by the high redshift data. The star forming galaxies tend to have the reddest NIR colors, but we can not say for certain that dust extinction is the main culprit. Most of the GDDS galaxies have total extinction values A$_{\rm v} <1.5\,$mag. The extinction vectors here suggest the extinction is in excess of A$_{\rm v} >2\,$mag. What do the resolved colors of nearby galaxies tell us about dust extinction in the NIR?  

\vspace{1cm}

\section{Sample Selection}\label{s:data}

Table~\ref{tab:summary} summarizes our sample of 68 spatially resolved nearby galaxies, with observations at 1.25\,\micron~($J$-band), 3.6\,\micron, 4.5\,\micron, 5.6\,\micron, 8.0\,\micron ~and 24\,\micron. All data come from publicly available sources, namely the \textit{Spitzer} Infrared Nearby Galaxies Survey (SINGS; \citealt{ken03}) 
and the Two Micron All Sky Survey (2MASS) Large Galaxy Atlas survey \citep[hereafter referred to as the 2MASS LGA]{jar03}.
We used SINGS data release {\tt 2070410\_enhanced\_v14}. Our data is comprised of all galaxies that were observed by
both surveys that meet our minimum signal-to-noise criterion (described below). As a result we excluded SINGS galaxies DDO\,053, Holmberg~I, Holmberg~IX, M81\,Dwarf~A and M81\,Dwarf~B, which do not have NIR coverage in the 2MASS~LGA. In addition, two SINGS galaxies, DDO\,154 and DDO\,165, did not contain any pixels which satisfied our signal-to-noise criterion, so only two objects were omitted for our analysis because of low signal-to-noise.  

\begin{deluxetable*}{cccccc}
\tablecaption{\small Object List}
\tablecolumns{6}
\tablewidth{0pc}
\tabletypesize{\small}
\tablehead{
	\colhead{Galaxy Name} 		&
	\colhead{Optical Morphology} 	&
	\colhead{Distance (Mpc)}	&
	\colhead{Metallicity [O/H]} &
	\colhead{$^a$AGN ?}	 &
	\colhead{$^b$H$\alpha$ ?} 	
}
\startdata 
     Ho\,ii  &        Im  &    3.5   &    0.0   &  0 &  1 \\ 
   IC\,2574  &      SABm  &    3.5   &    8.1   &  0 &  1 \\ 
   IC\,4710  &       SBm  &    8.5   &    0.0   &  0 &  1 \\ 
    MRK\,33  &        Im  &   21.7   &    0.0   &  0 &  1 \\ 
  NGC\,0024  &       SAc  &    8.2   &    0.0   &  0 &  1 \\ 
  NGC\,0337  &       SBd  &   24.7   &    0.0   &  0 &  1 \\ 
  NGC\,0584  &        E4  &   27.6   &    0.0   &  0 &  1 \\ 
  NGC\,0628  &       SAc  &   11.4   &    9.1   &  0 &  1 \\ 
  NGC\,0855  &         E  &    9.6   &    0.0   &  0 &  0 \\ 
  NGC\,0925  &      SABd  &   10.1   &    8.7   &  0 &  1 \\ 
  NGC\,1097  &       SBb  &   16.9   &    9.0   &  0 &  1 \\ 
  NGC\,1266  &       SB0  &   31.3   &    0.0   &  1 &  1 \\ 
  NGC\,1291  &       SBa  &    9.7   &    0.0   &  1 &  1 \\ 
  NGC\,1316  &      SAB0  &   26.3   &    0.0   &  1 &  1 \\ 
  NGC\,1377  &        S0  &   24.4   &    0.0   &  0 &  1 \\ 
  NGC\,1404  &        E1  &   25.1   &    0.0   &  0 &  1 \\ 
  NGC\,1482  &       SA0  &   22.0   &    0.0   &  0 &  1 \\ 
  NGC\,1512  &      SBab  &   10.4   &    0.0   &  0 &  1 \\ 
  NGC\,1566  &     SABbc  &   18.0   &    9.0   &  1 &  1 \\ 
  NGC\,1705  &        Am  &    5.8   &    0.0   &  0 &  1 \\ 
  NGC\,2403  &     SABcd  &    3.5   &    8.8   &  0 &  0 \\ 
  NGC\,2798  &       SBa  &   24.7   &    0.0   &  0 &  1 \\ 
  NGC\,2841  &       SAb  &    9.8   &    9.3   &  1 &  1 \\ 
  NGC\,2915  &        I0  &    2.7   &    0.0   &  0 &  0 \\ 
  NGC\,2976  &       SAc  &    3.5   &    0.0   &  0 &  1 \\ 
  NGC\,3031  &      SAab  &    3.5   &    8.9   &  1 &  1 \\ 
  NGC\,3034  &        IO  &    3.5   &    0.0   &  0 &  1 \\ 
  NGC\,3049  &      SBab  &   19.6   &    8.9   &  0 &  1 \\ 
  NGC\,3184  &     SABcd  &    8.6   &    9.3   &  0 &  1 \\ 
  NGC\,3190  &      SAap  &   17.4   &    0.0   &  1 &  1 \\ 
  NGC\,3198  &       SBc  &    9.8   &    8.9   &  1 &  1 \\ 
  NGC\,3265  &         E  &   20.0   &    0.0   &  0 &  1 \\ 
  NGC\,3351  &       SBb  &    9.3   &    9.3   &  0 &  0 \\ 
  NGC\,3521  &     SABbc  &    9.0   &    9.1   &  0 &  1 \\ 
  NGC\,3621  &       Sad     &    6.2   &  9.0  &  1  &  1 \\
  NGC\,3627  &      SABb  &    8.9   &    9.3   &  1 &  1 \\ 
  NGC\,3773  &       SA0  &   12.9   &    0.0   &  0 &  1 \\ 
  NGC\,3938  &       SAc  &   12.2   &    0.0   &  0 &  1 \\ 
  NGC\,4125  &       E6p  &   21.4   &    0.0   &  1 &  1 \\ 
  NGC\,4236  &      SBdm  &    3.5   &    0.0   &  0 &  0 \\ 
  NGC\,4254  &       SAc  &   20.0   &    9.2   &  0 &  1 \\ 
  NGC\,4321  &     SABbc  &   20.0   &    9.3   &  0 &  1 \\ 
  NGC\,4450  &      SAab  &   20.0   &    0.0   &  1 &  1 \\ 
  NGC\,4536  &     SABbc  &   25.0   &    8.9   &  0 &  1 \\ 
  NGC\,4552  &         E  &    4.5   &    0.0   &  0 &  0 \\ 
  NGC\,4559  &     SABcd  &   11.6   &    8.7   &  0 &  1 \\ 
  NGC\,4569  &     SABab  &   20.0   &    9.3   &  1 &  0 \\ 
  NGC\,4579  &      SABb  &   20.0   &    9.3   &  1 &  1 \\ 
  NGC\,4594  &       SAa  &   13.7   &    0.0   &  1 &  1 \\ 
  NGC\,4625  &     SABmp  &    9.5   &    0.0   &  0 &  0 \\ 
  NGC\,4631  &       SBd  &    9.0   &    0.0   &  0 &  1 \\ 
  NGC\,4725  &     SABab  &   17.1   &    9.3   &  1 &  1 \\ 
  NGC\,4736  &      SAab  &    5.3   &    9.0   &  1 &  0 \\ 
  NGC\,4826  &      SAab  &    5.6   &    9.4   &  1 &  1 \\ 
  NGC\,5033  &       SAc  &   13.3   &    9.1   &  1 &  0 \\ 
  NGC\,5055  &      SAbc  &    8.2   &    9.3   &  1 &  1 \\ 
  NGC\,5194  &     SABbc  &    8.2   &    9.3   &  1 &  1 \\ 
  NGC\,5195  &      SB0p  &    8.2   &    0.0   &  1 &  0 \\ 
  NGC\,5408  &       IBm  &    4.5   &    8.0   &  0 &  0 \\ 
  NGC\,5474  &      SAcd  &    6.9   &    0.0   &  0 &  1 \\ 
  NGC\,5713  &    SABbcp  &   26.6   &    0.0   &  0 &  1 \\ 
  NGC\,5866  &        S0  &   12.5   &    0.0   &  1 &  1 \\ 
  NGC\,6822  &       IBm  &    0.6   &    8.1   &  0 &  1 \\ 
  NGC\,6946  &     SABcd  &    5.5   &    9.1   &  0 &  1 \\ 
  NGC\,7331  &       SAb  &   15.7   &    9.1   &  1 &  1 \\ 
  NGC\,7552  &       SAc  &   22.3   &    9.0   &  1 &  1 \\ 
  NGC\,7793  &       SAd  &    3.2   &    8.6   &  0 &  1 \\ 
    TOL\,89  &      SBdm  &   15.0   &    0.0   &  0 &  1 \\ 
 
\enddata
\label{tab:summary}
\tablenotetext{a}{Shows spectral signatures of an active galactic nuclei \citep{dal06}.}
\tablenotetext{b}{Indicates H$\alpha$ observations}
\end{deluxetable*}
In addition to the infrared data contained in the SINGS and 2MASS LGA data releases, 56 galaxies had publicly-available H$\alpha$  images, obtained from telescopes at \textit{Kitt Peak National Observatory} (\textit{KPNO}) or \textit{Cerro Tololo Inter-American Observatory} (\textit{CTIO}). We used the continuum subtracted H$\alpha$ images contained within the SINGS data release, applying a small correction for a residual background pedestal where needed.

Because nuclear activity may contribute significantly to NIR and MIR flux,
we took careful consideration of which systems hosted active galactic nuclei.
Galaxies were differentiated based on nuclear activity using the results from \citet{dal06}, who used optical line diagnostics to identify which SINGS nuclei are either starburst dominated or require additional powering from an AGN, using the Liner/Starburst boundary of \ion{N}{2} $\lambda 6583 $/H$\alpha \sim 0.6$. This identification is 
included in Table~\ref{tab:summary}.

\section{Methods}\label{s:pp}

As we have noted in \S\,\ref{s:mot} and illustrated in Figure~\ref{fig:intcolors}, the scatter in the NIR colors of high-redshift star forming galaxies is greater than that seen nearby (where SFRs are lower). How does this small scatter manifest itself on smaller spatial scales? Our methodology is to group the pixels composing the local galaxy sample into regions based on different color-cuts motivated by both the spread in the data and the extinction law from \citet{Charlot:2000p1497} as indicated by the colored parallelograms shown in Figure~\ref{fig:intcolors}. The upper and lower limits of the boxes are defined as 0.1 dex above and below the extinction vector of \citet{Charlot:2000p1497}. The regions are separated in steps of increasing log$_{10}$~I$_{3.6}$/I$_{1.25} = 0.2$\,dex (or 0.5 mags in $L-J$ color) from  log$_{10}$~I$_{3.6}$/I$_{1.25} = -0.6$. We summarize the properties of the NIR color-selected regions in Table~\ref{tab:col}.

\begin{deluxetable}{ccc}
\tablecaption{\small Color key for Near-infrared color criteria}
\tablecolumns{3}
\tablewidth{0pc}
\tabletypesize{\small}
\tablehead{
	\colhead{Color Key} 		&
	\colhead{$^a\log_{10}$ I$_{3.6}$/I$_{1.25}$ }	 	&
	\colhead{$^b\log_{10}$ I$_{4.5}$/I$_{3.6}$ }	 	
}
\startdata 
Red 		&  -0.6 to -0.4	&  -0.19 \\
Orange	&  -0.4 to -0.2	&  -0.17 \\
Yellow	& -0.2 to 0.0	& -0.14 \\ 
Green	& 0.0 to 0.2	& -0.12 \\
Blue		& $>$0.2		& -0.09 \\	
Purple	& Manually selected  \\
\enddata
\label{tab:col}
\tablenotetext{a}{Upper and lower limits.}
\tablenotetext{b}{Value in the center of the parallelogram shown in Figure~\ref{fig:intcolors}.}
\end{deluxetable}

\subsection{Image preparation}

Prior to analyzing the pixel-by-pixel colors of our galaxy sample a number of steps were
taken to register images to a common pixel scale and common point-spread-function 
(PSF), clean out contaminants and ensure that each pixel has a signal-to-noise ($S/N$) ratio of at least 15. Each step is outlined below in more detail.

\begin{enumerate} 
\item All images were re-sampled to the IRAC plate scale of 1.22\,\arcsec/pixel using the {\tt remap} tool in the \nobreak{\tt WCStools} library \citep{min02}. 
\item The counts in each pixel were converted to intensity units of MJy/sr. 
\item Images were registered to sub-pixel accuracy using unsaturated bright stars and 
500 by 500 pixel ($\sim$10.2\,\arcmin~by $\sim$10.2\,\arcmin) subsets were cut from the registered master frames.
\item Saturated regions and large galaxy neighbours on each image were manually masked.
\item All images were convolved with a gaussian kernel to match the seeing-limited $\sim$6\,\arcsec~beam of the 24\,\micron~MIPS image point-spread function. We assumed a median FWHM of 1.0\,\arcsec, 2.8\,\arcsec, 1.7\,\arcsec, 1.7\,\arcsec, 1.9\,\arcsec, 2.0\,\arcsec~and 6.0\,\arcsec~for the 1.25, 3.6, 4.5, 5.6, 8.0 and 24\,\micron~images, respectively. (Pixel-by-pixel image analysis requires that each pixel value be a good representation of the true intensity in the area of galaxy contained by the pixel. By convolving our images, we ensured that the distribution of intensities were matched as they were spread across the pixels by varying point-spread functions.) 
\item Bright foreground stars and small background galaxies were automatically removed.  Foreground stars were removed in 5 pixel (6.1\arcsec) radius apertures using the the 2MASS point source catalog. Unfortunately many compact \ion{H}{2} regions are falsely identified as point sources in the 2MASS catalog, so we attempted to identify these sources beforehand through their mid-infrared colors as in \citet{MunozMateos:2009p537}. We also did not remove any point sources with 24\,\micron~intensity $>0.7$\,MJy. These two criteria result in $\sim$30\% catalog point source removal, with some contaminants remaining, but we chose this approach to preserve compact \ion{H}{2} regions and compact nuclear sources such as AGN. In addition, faint background galaxies have the potential to show up as spurious objects in our analysis.  We have inspected each image carefully to note cases where contamination remains important after the steps just described, and will explicitly note these cases below.
\item We used a median adaptive smoothing algorithm \citep{Zibetti:2009p710} on the 1.25, 3.6, 4.5 \micron~images so that $S/N > 15$ is obtained in each pixel. This final step is non-standard and will be described below. 
\end{enumerate} 

\subsection{Adaptive smoothing}

Much of our analysis is based on an investigation of the flux ratios of individual pixels. This required the construction of data frames obtained by dividing 1.25\,\micron, 3.6\,\micron, and 4.5\,\micron~images against each other. To minimize the noise in the resulting divided images we used the {\tt Adaptsmooth} software \citep{Zibetti:2009p710} to median smooth the images so that a reasonably high signal-to-noise ($S/N$) threshold is first met. This trades resolution at low signal-to-noise levels for robustness in the final flux ratios (see the discussion in Zibetti et al. 2009 for details). 

Our goal with {\tt Adaptsmooth} was to obtain input data frames where $S/N > 15$ in each pixel. 
{\tt Adaptsmooth} works by taking the median intensity over a smoothing aperture with a radius defined by the size needed to meet the $S/N > 15$ criterion. If the smoothing radius required is greater than 12 pixels, the algorithm is declared to have failed for that pixel, and the pixel is excluded from subsequent analysis. All results reported in this analysis in which colors were computed in individual pixels use the smoothed then convolved images. For analysis involving measurements of the average intensity over many pixels
adaptsmoothed images are used to perform pixel selection (which
would be impossible otherwise), whereas unsmoothed images are used to
derive pixel intensities to be averaged. In fact, averaging over many
pixels makes the adaptsmoothing step superfluous. Moreover, this
approach allows us to retain the most accurate flux information in
each pixel and avoid possible biases (although these biases are modest, as discussed in
\citealt{Zibetti:2009p710}).

\section{Results}\label{s:results}

\begin{figure*}[tbp]
\begin{center}
\includegraphics[width=7in]{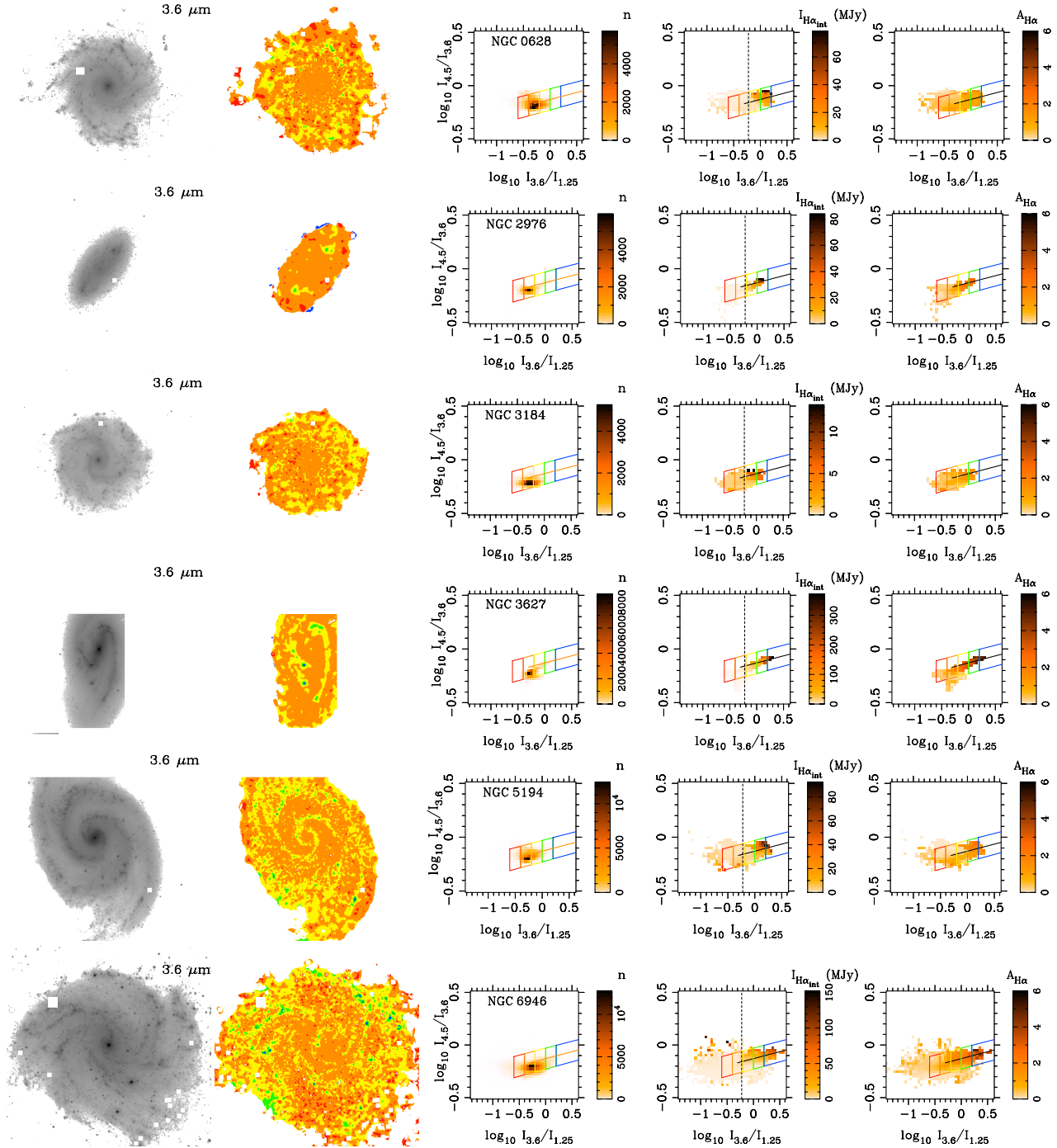}
\caption[Postage stamps and NIR pixel color-color diagrams of SINGS galaxies]{\small Images at IRAC 3.6\,\micron~and the spatial locations of the NIR color selected pixels (defined in Table~\ref{tab:col}) are shown for examples of normal star-forming galaxies in our sample: NGC\,0628, NGC\,2976, NGC\,3184, NGC\,3627, NGC\,5194 and NGC\,6946. The three rightmost panels plot the pixel NIR colors with the intensity levels defined as the number of pixels at each color bin in the first of these three panels. NIR color selection criteria is shown in colored boxes. The last two panels show the \textit{median} intrinsic H$\alpha$ intensity (and consequently star formation rate density) and the median $A(H\alpha)$ in mag for all the pixels in that NIR color bin. Figures for all galaxies can be found in the online edition.}
\label{fig:pixels}
\end{center}
\end{figure*}

The near-infrared colors in the individual pixels of the 68 galaxies in our sample were measured by performing image division with the median smoothed (and signal-to-noise optimized) then convolved images (as described in \S\,\ref{s:pp}). The resulting pixel colors in the NIR color space are shown for 6 example galaxies from our sample in Figure~\ref{fig:pixels}. Results for all galaxies can be found on the online edition of this article. We find 10 galaxies to have atypical pixel distributions and these are discussed later in \S\,\ref{s:odd}. In the leftmost column, 10.2\arcmin$\times$10.2\arcmin~images of the galaxies are shown at IRAC 3.6\,\micron. Regions with saturated objects, foreground stars or neighbouring galaxies are removed. In the column to the right of the image column, we indicate the spatial location of the NIR color-selected pixels. The colors are keyed according to the NIR color-selection, shown in Figure~\ref{fig:intcolors} and defined in Table~\ref{tab:col}.

The three rightmost columns show binned color-color diagrams with the intensity of cells keyed to various quantities. The intensity in the first of these three is keyed to the total number of pixels in that color bin. The scatter in pixel NIR colors is larger than for the integrated colors of the SINGS galaxies shown in Figure~\ref{fig:intcolors}, with some pixels containing intensity ratios at  I$_{3.6}$/I$_{1.25} > 1$, larger than any of the integrated colors. As with the integrated colors, the majority of pixels in the galaxies are found in the orange region, that is 0.1\,dex (or 0.25 mags) above and below our typical color for the sample of log$_{10}$ I$_{3.6}$/I$_{1.25}$, log$_{10}$ I$_{4.5}$/I$_{3.6}$ = -0.30, -0.17, or (1.25]-[3.6]$_\mathrm{AB}$ = -0.75 and [3.6]-[4.5]$_\mathrm{AB}$ =  -0.425). For the spatially resolved galaxy sample, we find the majority of pixels are found at log$_{10}$ I$_{3.6}$/I$_{1.25} = -0.30\pm0.07$ and log$_{10}$ I$_{4.5}$/I$_{3.6} = -0.19\pm0.02$, consistent with the colors we visually selected from the integrated colors to represent the colors of evolved stellar populations. A smaller subset of pixels tend to scatter along the extinction vector defined by $\lambda^{-0.7}$ from empirical modeling by \citet{Charlot:2000p1497}, shown to be a good match to starburst galaxies from \citet{Calzetti:1994p1499}. 

Figure~\ref{fig:pixels} allows us to map the color-selected pixels into the spatial domain of the galaxy. The majority of pixels with NIR colors (keyed as orange regions) described by stellar emission are found throughout the galaxy. Regions with redder colors (identified as yellow then green then blue) follow the spiral structure and regions of star formation in the galaxy. A degeneracy between extinction and emission occurs at the brightest knots of star formation, where the NIR colors are their reddest. If these regions correspond to regions of high dust extinction, then the red NIR colors may be the result of dust extinction in which the light at 3.6\,\micron~is attenuated less than the 1.25\,\micron~emission. On the other hand, additional emission can also  explain the excess continuum emission at $2-5$\,\micron.

Fortunately, for 56 of the galaxies, continuum subtracted H$\alpha$ images can be combined with 24\,\micron~images to create two composite images that represent the intrinsic H$\alpha$ emission (corrected for attenuation due to dust extinction) and the dust extinction at H$\alpha$ in each pixel. The intrinsic H$\alpha$ line emission luminosity, and consequently the star formation rate, of normal star forming galaxies and \ion{H}{2} regions can be estimated empirically as a linear superposition of the observed H$\alpha$ and 24\,\micron~flux \citep{ken07,cal07}. Recent work \citep{ken09} shows that for a sample of galaxies, including the SINGS galaxies, the dust corrected, intrinsic H$\alpha_{int}$ luminosity is related to the observed H$\alpha_{obs}$ and 24\,\micron~luminosity by:

\begin{center}
\begin{equation}
L(\mathrm{H}\alpha_{int}) = L(\mathrm{H}\alpha_{obs}) + 0.02 \times L(24\,\micron)
\end{equation}
\end{center}
\vspace{0.2cm}

\noindent with the galaxies showing a rms dispersion of 0.119 dex in the fit.  From this equation, one can immediately get I$_{\mathrm{H}\alpha,{int}}$ shown in Equation 2. Also, from L(H$_{\alpha,int}$) one can obtain the SFR by means of a simple multiplicative factor (e.g. \citealt{Kennicutt:1998p693}). Hence, I$_{\mathrm{H}\alpha,{int}}$, is in fact a direct estimate of the local star formation rate density (or surface brightness).

\begin{center}
\begin{equation}\label{eq:halpha}
I_{\mathrm{H}\alpha,{int}} =  I_{\mathrm{H}\alpha,{obs}} + 0.02 \times \frac{24.0}{0.656} \times I_{24\,\micron}
\end{equation}
\end{center}
\vspace{0.2cm}

\noindent Similarly, the average extinction in each pixel can be derived using the intrinsic H$\alpha$ intensity, $I_{\mathrm{H}\alpha,{int}}$, defined above and the observed H$\alpha$ intensity, $I_{\mathrm{H}\alpha,{obs}}$ . The extinction, $A(H\alpha)$ in mag, is given as:

\begin{center}
\begin{equation}\label{eq:dust}
A(H\alpha) = 2.5 \log_{10} \left[1 +  0.02\times(24.0/0.656) \frac{I_{24\,\micron}}{I_{H\alpha,obs}} \right]
\end{equation}
\end{center}
\vspace{0.2cm}

The two panels on the right of Figure~\ref{fig:pixels} show the median intrinsic H$\alpha$ intensity, $I_{{\rm H}\alpha,int}$, and the median dust extinction, $A(H\alpha)$, in the pixels corresponding to the given NIR colors. By spatially resolving the galaxy colors, we can see a trend emerge in the NIR color-color plane in the pixel plots shown in Figure~\ref{fig:pixels}. The pixels which scatter along the extinction line to the redder color-selected regions also tend to have the highest intrinsic H$\alpha$ intensities. Regions with higher intensity star formation show redder NIR colors. The dust extinction, in the far right panel, shows less variation, and tends to be $A(H\alpha) = 0-2$~mag. There is no strong trend of increasingly red NIR colors with increasing dust extinction. We now attempt to quantify this by measuring the average surface brightness in increasingly red NIR color regions.

\begin{figure}[tbp]
\begin{center}
\plottwo{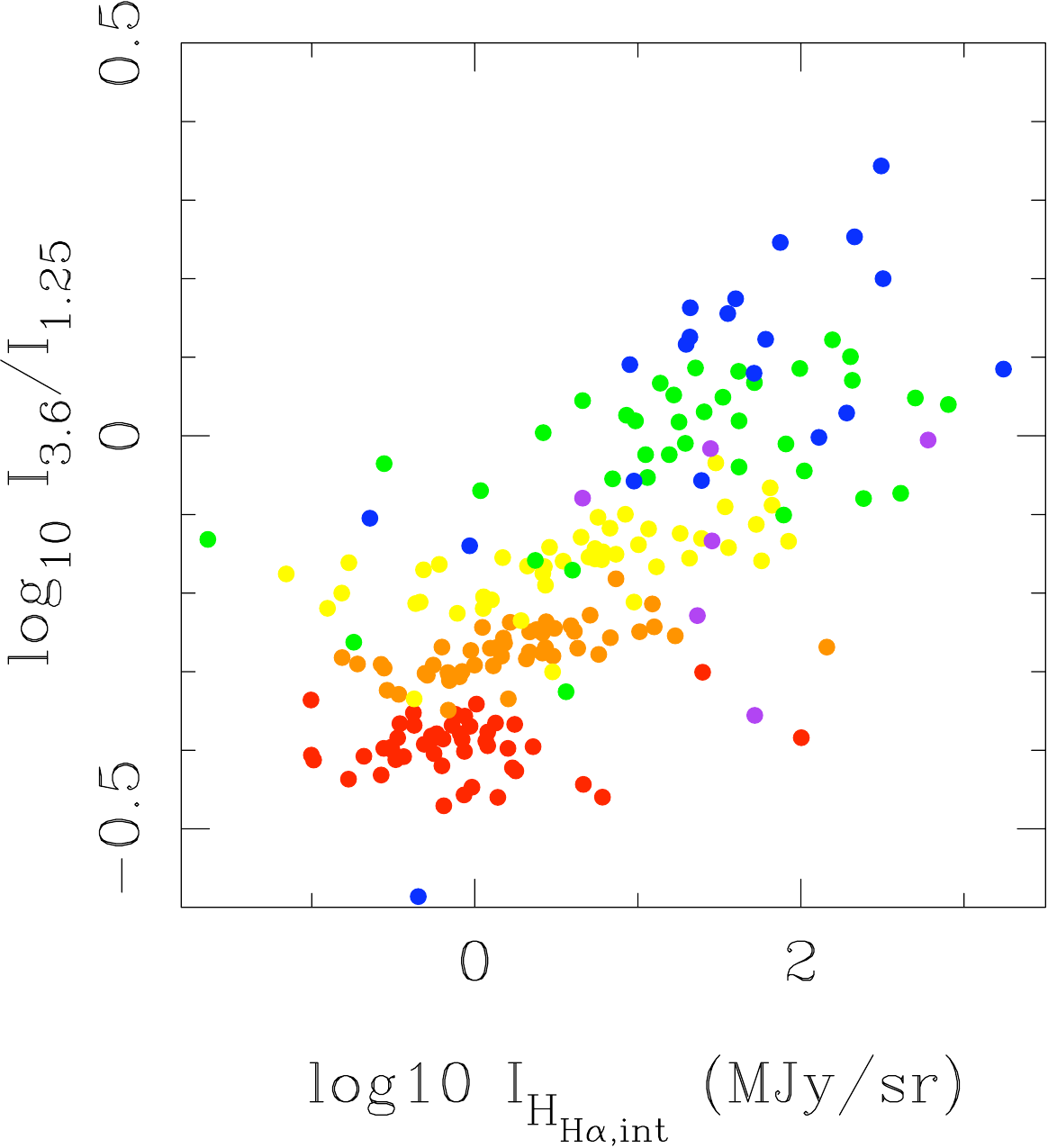}{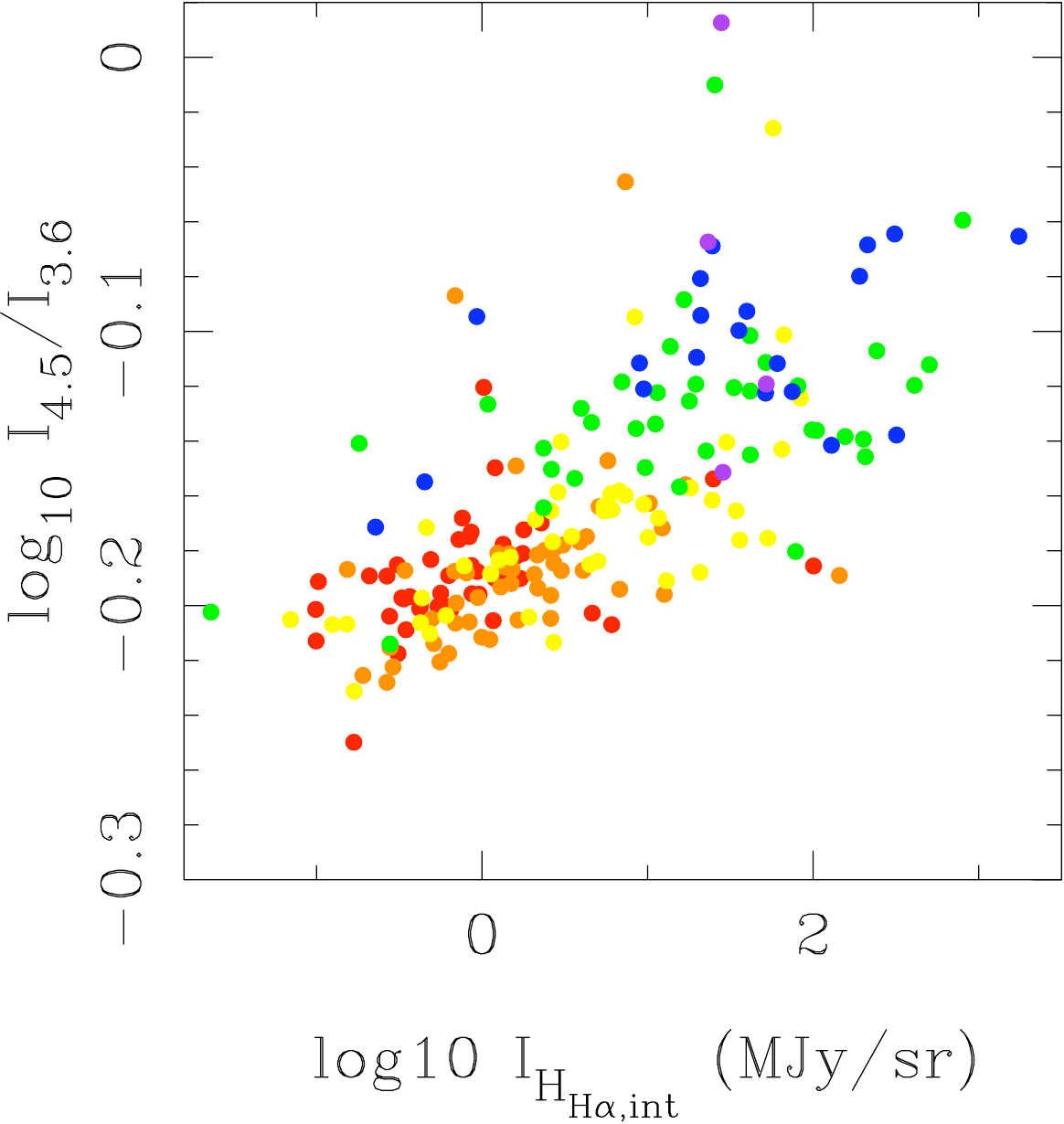}
\plottwo{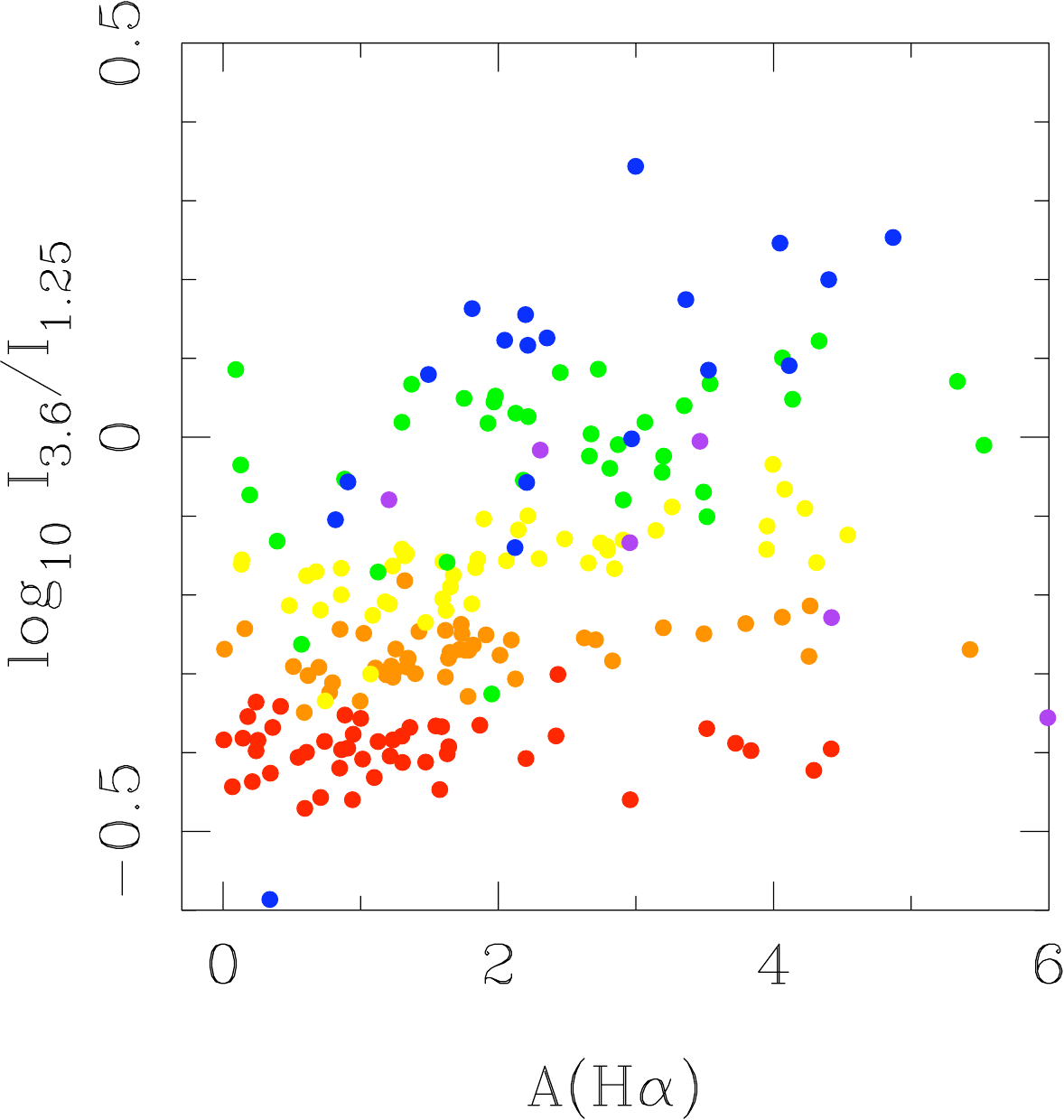}{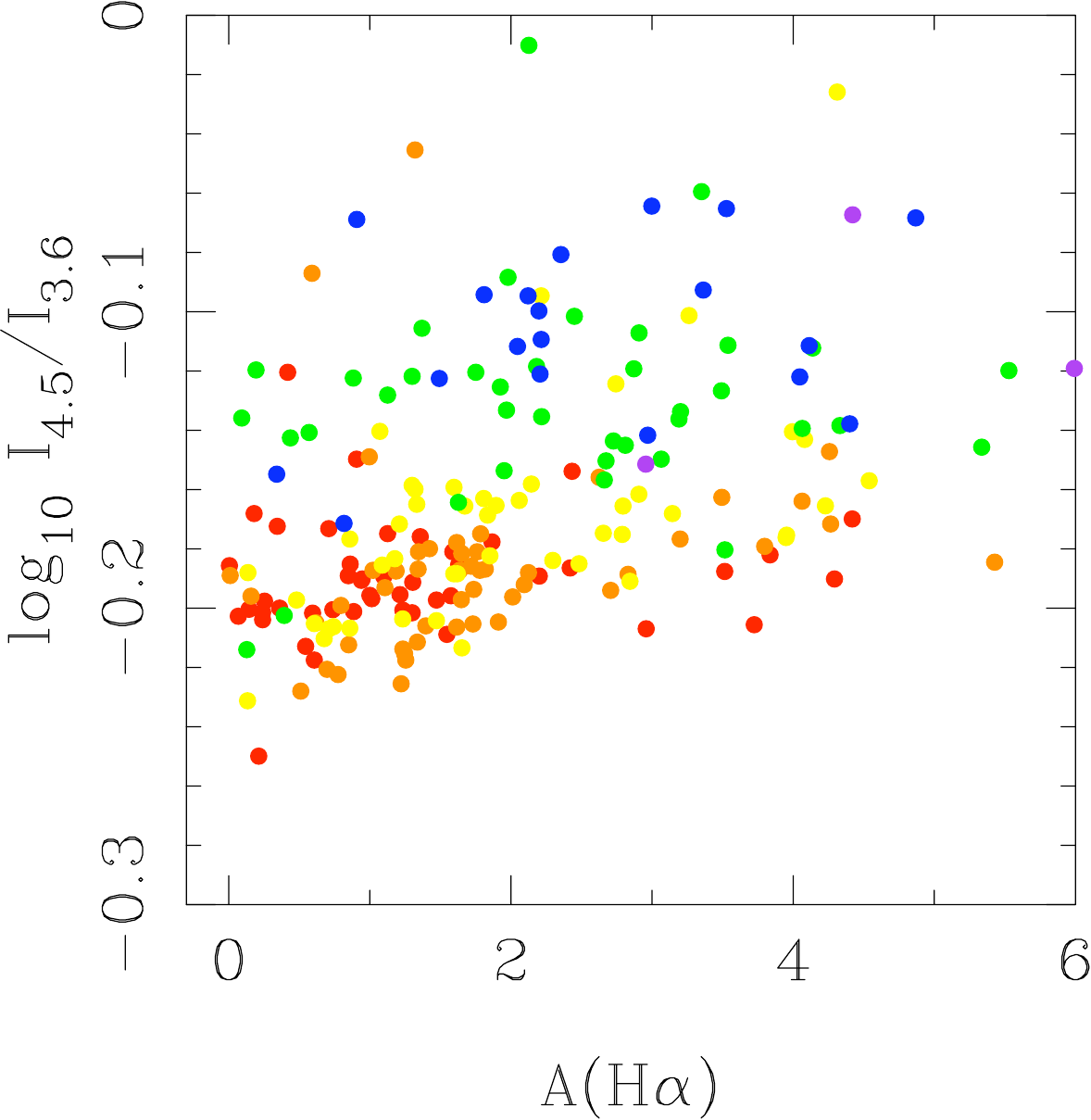}
\caption[NIR colors and dust extinction]{The average I$_{3.6}$/I$_{1.25}$ (left column) and I$_{4.5}$/I$_{3.6}$ (right column) colors in each of the 5 regions defined in Figure~\ref{fig:pixels} as a function of the average dust-corrected H$\alpha$ intensity (top panels) and dust extinction at H$\alpha$ (bottom panels). The colors are keyed to the NIR color selected regions defined in\S\,\ref{s:pp} and described in Table~\ref{tab:col}. The purple dots are regions with anomalous NIR color sequences as discussed in \S\,\ref{s:odd}. By design, the colors should change from red to blue with increasing I$_{3.6}$/I$_{1.25}$ and I$_{4.5}$/I$_{3.6}$ ratios. The dust extinction is scattered with values ranging from 0 to 6 mags of extinction in all color-selected regions. There is a stronger correlation between redder NIR colors with the average dust-corrected H$\alpha$ intensity than average dust extinction.} 
\label{fig:MLvsAha}
\end{center}
\end{figure}

\subsection{Average surface brightness measurements in NIR color-selected regions}\label{s:sb}

We measured the average NIR colors and intensity (i.e. surface brightness) for each galaxy in each of the NIR color selected regions marking 0.1\,dex in color space above and below the extinction curve of $\lambda^{-0.7}$ from \citet{Charlot:2000p1497} and defined in Table~\ref{tab:col}. The regions are mapped in the spatial domain and in the NIR color plane in the second and third columns of Figure~\ref{fig:pixels}. Measurements are made at all wavelengths in this analysis (0.6563, 1.25, 3.6, 4.5, 5.6, 8.0 and 24\,\micron) as well as to the intrinsic H$\alpha$ image defined by Equation~\ref{eq:halpha} and the $A(H\alpha)$ dust extinction image defined by Equation~\ref{eq:dust}.

We begin by looking at the NIR color trends. A trend of increasing I$_{3.6}$/I$_{1.25}$ and/or  I$_{4.5}$/I$_{3.6}$ colors with increasing dust extinction would suggest that red NIR colors are due to dust, or alternatively, a trend of increasingly red NIR colors with the average dust-corrected H$\alpha$ surface brightness would suggest that the NIR colors are instead due to dust emission. Because of the likely coincidence of both high dust extinction and strong H$\alpha$ emission in regions of star formation, it is not easy to dissociate these parameters. Our attempt to do so is shown in Figure~\ref{fig:MLvsAha}. Here, the NIR colors in the color-selected regions are compared to the average dust-corrected H$\alpha$ emission and dust extinction. The top row shows the NIR colors (I$_{3.6}$/I$_{1.25}$ and I$_{4.5}$/I$_{3.6}$ from left to right) both increase with increasing H$\alpha$ emission, as they do with dust extinction in the bottom rows. The data points are keyed to the color selection regions. As expected (by design) the regions with higher NIR ratios are from the redder color-selected regions (blue, green and yellow points). 

In all cases, there is a trend of increasing NIR colors with both H$\alpha$ emission and dust extinction. We find the correlation is stronger between the NIR colors and the H$\alpha$ emission than the dust extinction. Quantified by \textit{Pearson's correlation coefficient}, the NIR intensity ratios, I$_{3.6}$/I$_{1.25}$ and I$_{4.5}$/I$_{3.6}$, correlate to the intrinsic H$\alpha$ surface brightness by $\rho = 0.38$ and 0.36, respectively. We compare how the NIR intensity ratios correlate with the dust attenuation, quantified as I$_{H\alpha_{int}}/I_{H\alpha_{obs}}$ (to compare linear quantities), and find that the correlation is weaker, but still significant at $\rho = 0.11$ and 0.16 for I$_{3.6}$/I$_{1.25}$ and I$_{4.5}$/I$_{3.6}$ ratios respectively. \textit{The stronger trend with H$\alpha$ emission suggests the NIR colors are due to dust emission rather than extinction. The weaker trend for the reddest color-selected regions to have a higher median extinction value is likely a consequence of the higher star formation and associated dust content in these regions, resulting in higher dust extinction.}

\begin{figure*}[htbp]
\begin{center}
\includegraphics[width=7.in]{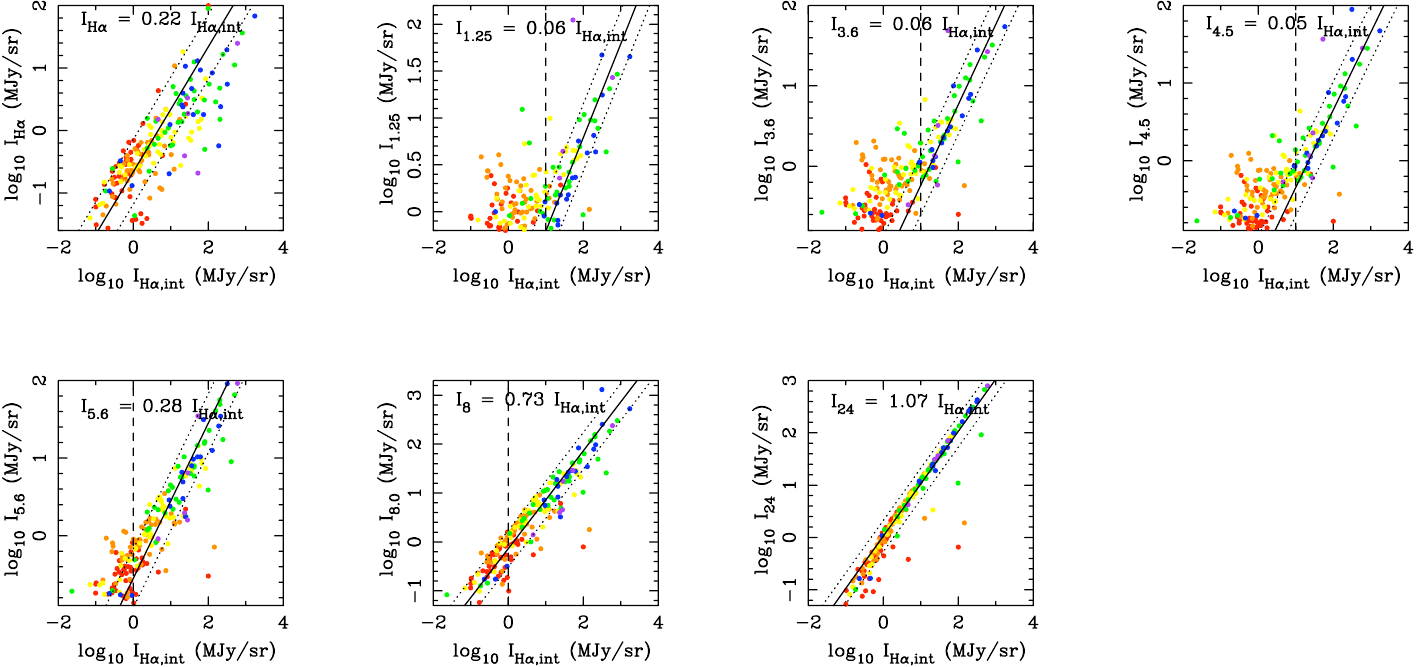}
\caption[Linear relationship between the NIR and MIR intensities and the intrinsic H$\alpha$ intensity]{Surface brightness measurements at H$\alpha$, 1.25, 3.6, 4.5, 5.6, 8.0 and 24. \micron~as a function of the intrinsic H$\alpha$ intensity (a direct measure of the average star formation rate per unit area). Colors are keyed to increasingly red NIR colors (as defined in Table~\ref{tab:col}). Linear fits with zero offsets are fit to each distribution, whose slope is given in each plot. The rms of the fit is shown by the dotted curves. Only points to the right of the dashed vertical line are used for the fit to minimize the contribution of emission due to underlying evolved stellar populations. All points are used for the H$\alpha$ and 24\,\micron~fits as the H$\alpha$ image is stellar subtracted and the stellar contribution at 24\,\micron~image is assumed be negligible. The linear fit to each relation (ie. the intensity ratio relative to I$_{\mathrm{H}\alpha,int}$) is tabulated in Table~\ref{tab:coefs}. Galaxies with anomalous NIR colors (discussed in \S\,\ref{s:odd}) whose pixels correspond to compact regions of star formation are shown as purple dots.  } 
\label{fig:sbtosfr}
\end{center}
\end{figure*}

The intensity in each region is compared for each observed waveband to the intrinsic H$\alpha$ intensity (a direct measure of the average star formation rate) in Figure~\ref{fig:sbtosfr}. There is clear correlation between the intensity at 5.6, 8.0 and 24\,\micron~and the intrinsic H$\alpha$ intensity. These bands are dominated by PAH emission. Also evident is a correlation between emission at 1.25, 3.6 and 4.5\,\micron~and the intrinsic H$\alpha$ emission, although here the trends are more nuanced. In regions of high intrinsic H$\alpha$  intensities (and high star formation rates), the emission from the young star-forming environment dominates over the emission of the longer lived population. But at lower H$\alpha$ intensities ($\lesssim$10\,MJy/sr), the emission at 3.6, 4.5\,\micron~and particularly at 1.25\,\micron~is scattered in this plane, indicating that in these regions the NIR emission from the star forming population is lower and the emission traces the evolved stellar populations which varies from galaxy to galaxy as expected from the different star formation histories. We calculate the intensity ratios at 1.25, 3.6, 4.5, 5.6, 8.0 and 24\,\micron~relative to I$_{\mathrm{H}\alpha,int}$ in Table~\ref{tab:coefs}. In the rightmost column, we give the range in possible values for the proportionality constant between I$_{\mathrm{H}\alpha,int}$ and intensity at other wavelengths within the rms of the fit.

\begin{figure*}[htbp]
\begin{center}
\includegraphics[width=7.in]{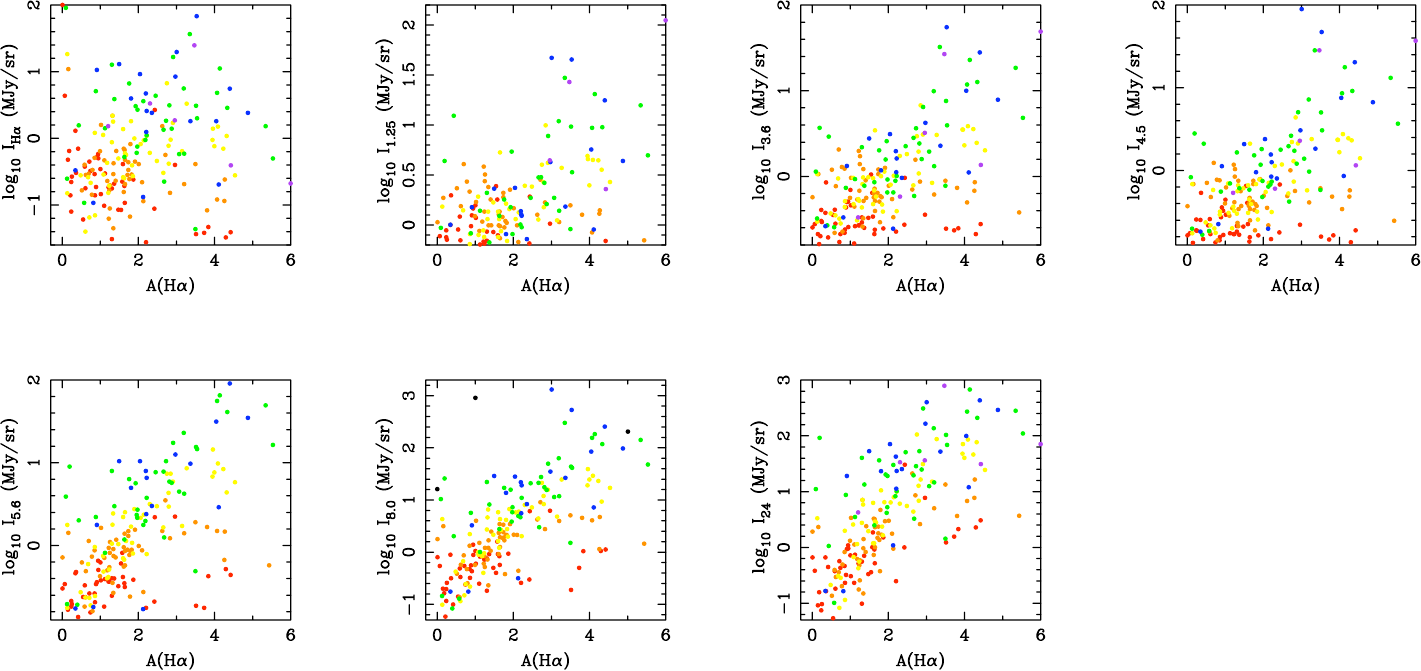}
\caption[Correlation between NIR and MIR intensities and dust extinction]{Surface brightness measurements at H$\alpha$ (0.656), J (1.25), 3.6, 4.5, 5.6, 8.0 and 24\,\micron~in relation to the average dust extinction derived from the ratio between the H$\alpha$ and 24\,\micron~intensities as described by Eq.\,\ref{eq:dust}. The different colored points are keyed to the regions of increasingly larger NIR colors as defined in \S\,\ref{s:pp} and plotted in Figure~\ref{fig:pixels}. The purple dots are regions with anomalous NIR color sequences as discussed in \S\,\ref{s:odd}.} 
\label{fig:sbtoAha}
\end{center}
\end{figure*}

These results support the conclusion that very red NIR colors are due to sources of NIR emission. But the possibility that extinction may also be important needs further exploration. In particular, the trend for increasingly red NIR colors falls along the dust law of \citet{Charlot:2000p1497}, which may suggest that the spread in NIR colors is at least partly due to dust extinction rather than any additional emission mechanism. Using the dust extinction image defined by Equation~\ref{eq:dust}, we can measure the average dust extinction in the NIR color-selected pixels. We further investigate any trends between average intensities at all observed wavebands and the average dust extinction in Figure~\ref{fig:sbtoAha}. Again, the data show a correlation with the dust extinction. At 5.6, 8.0 and 24\,\micron~larger extinction values correlate with higher intensities resulting from increased dust emission. These regions, as shown in Figure~\ref{fig:sbtosfr}, correspond to more intense star forming regions and as a result have the most extreme extinction values. However, we once again see that the correlation with the dust extinction is weaker in all wavebands compared with the correlation between intensities and intrinsic H$\alpha$ emission shown in Figure~\ref{fig:sbtosfr}. \textit{In conclusion, Figures~\ref{fig:MLvsAha} to \ref{fig:sbtoAha}, indicate that the 
the NIR color excess is due to emission and not dust extinction at wavelengths longer than 1\,\micron.}

\subsection{Galaxies with anomalous NIR color structures}\label{s:odd}

\begin{figure*}[tbp]
\begin{center}
\includegraphics[width=7in]{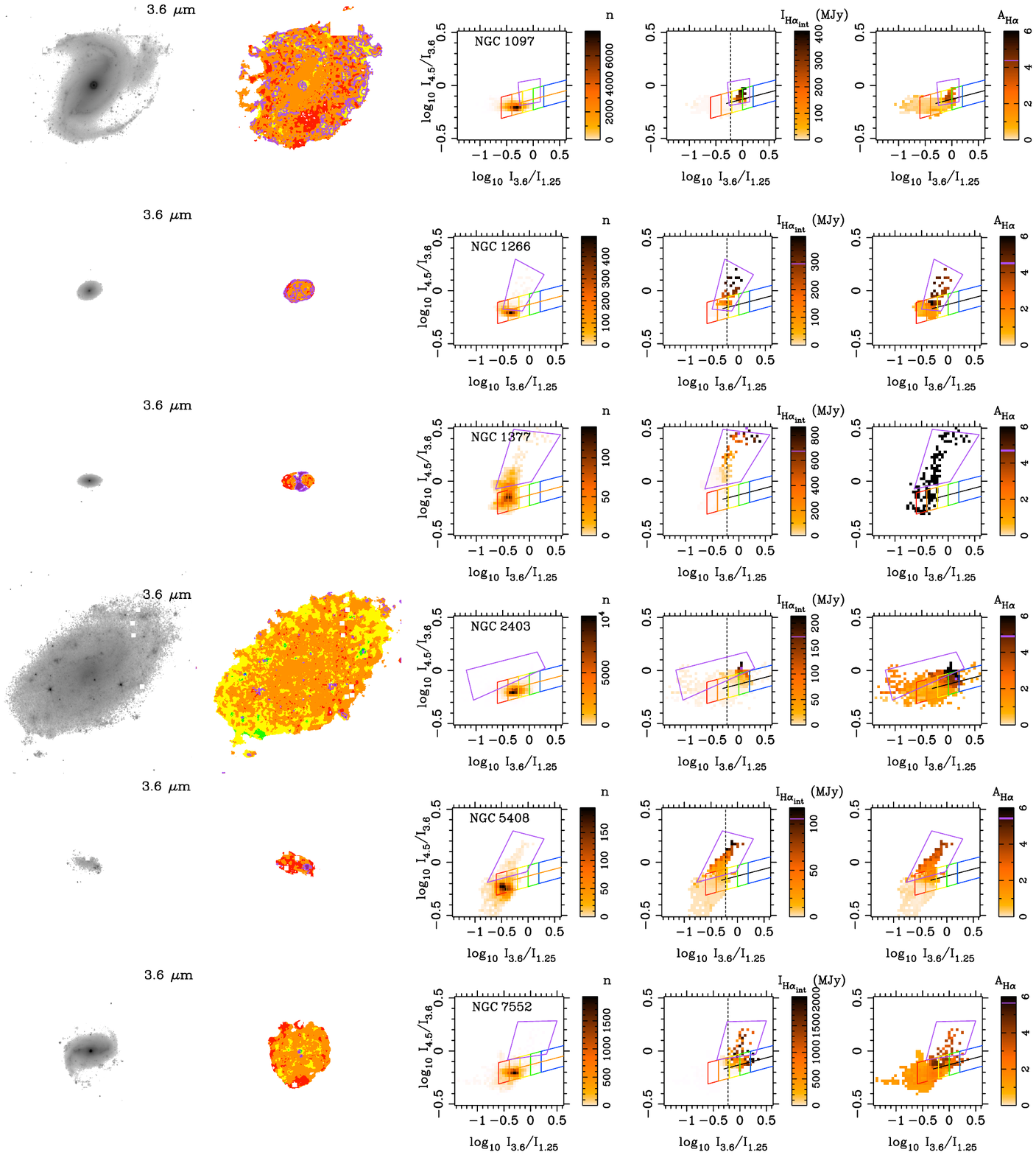}
\caption[Galaxies with compact regions of anomalous NIR color structures]{IRAC 3.6\,\micron~images for galaxies with anomalous NIR pixel colors are shown for (from top to bottom) NGC\,1097, NGC\,1266, NGC\,1377, NGC\,2403, NGC\,5408 and NGC\,7552. The second panel from the left shows the spatial location of the anomalous pixel colors (outlined in purple).  In general, these colors spatially correspond to bright compact regions of star formation. The left panel of the three pixel color panels shows the number of pixels in each color-color bin. The middle panel indicates the median value of the intrinsic H$\alpha$ intensity for all the pixels within that color-color bin. The left panel indicates the median dust extinction for the pixels.}
\label{fig:pixels_odd}
\end{center}
\end{figure*}

Not all galaxies show binned pixel distributions similar to the typical ones presented in Figure~\ref{fig:pixels}. Ten galaxies (15\%) show distributions with markedly different structures.  To study these, we manually selected regions in the NIR color space to identify the spatial location of anomalous populations. Their location in pixel space and in the spatial domain are shown in purple for a 6/10 of these galaxies in Figure~\ref{fig:pixels_odd} (the rest can be found in the online edition of Figure~\ref{fig:pixels}). For these 10 galaxies (Holmberg\,II, NGC\,1097, NGC\,1266, NGC\,1377, NGC\,2403, NGC\,3031, NGC\,4736, NGC\,5408, NGC\,7552, Tololo\,89) we find that the red colors are related to regions of clumpy star formation as indicated by compact 24\,\micron~emission. In most of these cases, the NIR colors follow a vector corresponding to higher I$_{4.5}$/I$_{3.6}$ ratios than the $\lambda^{-0.7}$ extinction law from \citet{Charlot:2000p1497}. For example, Holmberg\,II, NGC\,1266, NGC\,1377, NGC\,5408, NGC\,7552 and Tololo\,89 show a similar distribution of NIR colors. The increasingly red NIR colors correspond to regions of increasing star formation (4th panel in Figure~\ref{fig:pixels_odd}) and higher dust extinction (right panel). These regions correspond to compact regions with bright infrared emission. The emission is likely related to star formation; in many cases it corresponds to the brightest regions along a spiral arm or compact opaque star formation, as is the case for NGC\,1377 which is an opaque starburst galaxy whose dust emission is optically thick \citep{rou06}. In this galaxy, the emission is related to very young ($<1$\,Myr), highly obscured starburst activity.  These NIR colors are the result of either a different extinction law that manifests as an optically thick dust screen or perhaps a different dust emission law resulting from the high radiation fields produced by the starburst activity (in which the ratio of PAH emission and continuum emission from hot and warm dust is different than the typical spectrum found in normal star forming environments found in this study). Some additional possible mechanisms have recently put forth in \citet{smi09} to explain the redder I$_{4.5}$/I$_{3.6}$ ratios seen in nearby dwarf galaxies.

NGC\,3031 and NGC\,4736 display NIR color sequences that show increasing I$_{4.5}$/I$_{3.6}$ ratios  for decreasing I$_{3.6}$/I$_{1.25}$ ratios. Both of these galaxies show spectral indicators of harboring an AGN \citep{dal06}. 

We have included the points for these 10 galaxies in Figures~\ref{fig:sbtosfr}, \ref{fig:MLvsAha} \& \ref{fig:sbtoAha} as purple dots to compare the average intensities in these anomalous red colors sequences to the star formation rate intensity and dust extinction and find that in all cases the values agree with the measurements made for the standard NIR color-selected regions. 

\section{Discussion}\label{s:disc}

Spatially resolved pixel-by-pixel analysis of the SINGS galaxies yields a number of insights regarding the underlying stellar populations and their associated dust emission and extinction. In regions of negligible star formation, NIR colors occupy a small region of NIR color space, and we have shown that the majority of pixels tend to be located in this color space. However, younger populations, with higher H$\alpha$ and 24\,\micron~emission, tend to show colors that become increasingly red in the NIR, with both the intensity ratios of I$_{3.6}$/I$_{1.25}$ and I$_{4.5}$/I$_{3.6}$ increasing along a vector well matched by the $\lambda^{-0.7}$ extinction law from \citet{Charlot:2000p1497}. Motivated by this empirical trend in the data, we divided galaxies into different regions based on the NIR color of each individual pixel. These regions are designed to encompass 0.1\,dex (or 0.25 mags) above and below the $\lambda^{-0.7}$ extinction curve separated into steps of 0.2\,dex (0.5 mags) in $\log$\,I$_{3.6}$/I$_{1.25}$ color starting at $\log$\,I$_{3.6}$/I$_{1.25}=-0.6$. By doing so we isolate regions with different NIR emission properties and consequently different stellar populations, revealing the relatively linear nature of emission in the near- and mid-infrared.

\subsection{The linear scaling of the near/mid-infrared spectrum}

Figure \ref{fig:sbtosfr} shows that the average intensity of NIR color-selected regions is strongly correlated with with the average intrinsic H$\alpha$ intensity in that region for the majority of bands in this study. The 5.8, 8.0 and 24\,\micron~intensities scale linearly with this parameter, with scatter presumably mostly due the non-negligible contribution from the stellar population at 5.8 and 8.0\,\micron~and also due to differences in metallicity \citep{cal07} and possibly PAH fractions \citep{dra07}. Rather surprisingly, the shorter NIR wavelengths also show a linear correlation with the star formation rate intensity at sufficiently high intensities that the emission is dominated by emission from the dust associated with young stellar populations. We find that linear fits to the data with zero offsets can fit the data in all wavebands, but limit the fits to large star formation intensities (I$_\mathrm{SFR}>$10 MJy/sr) in the shorter bands and slightly lower star formation intensities (I$_\mathrm{SFR}>$1 MJy/sr) at 5.8 and 8.0\,\micron.

\begin{figure*}[htbp]
\begin{center}
\plotone{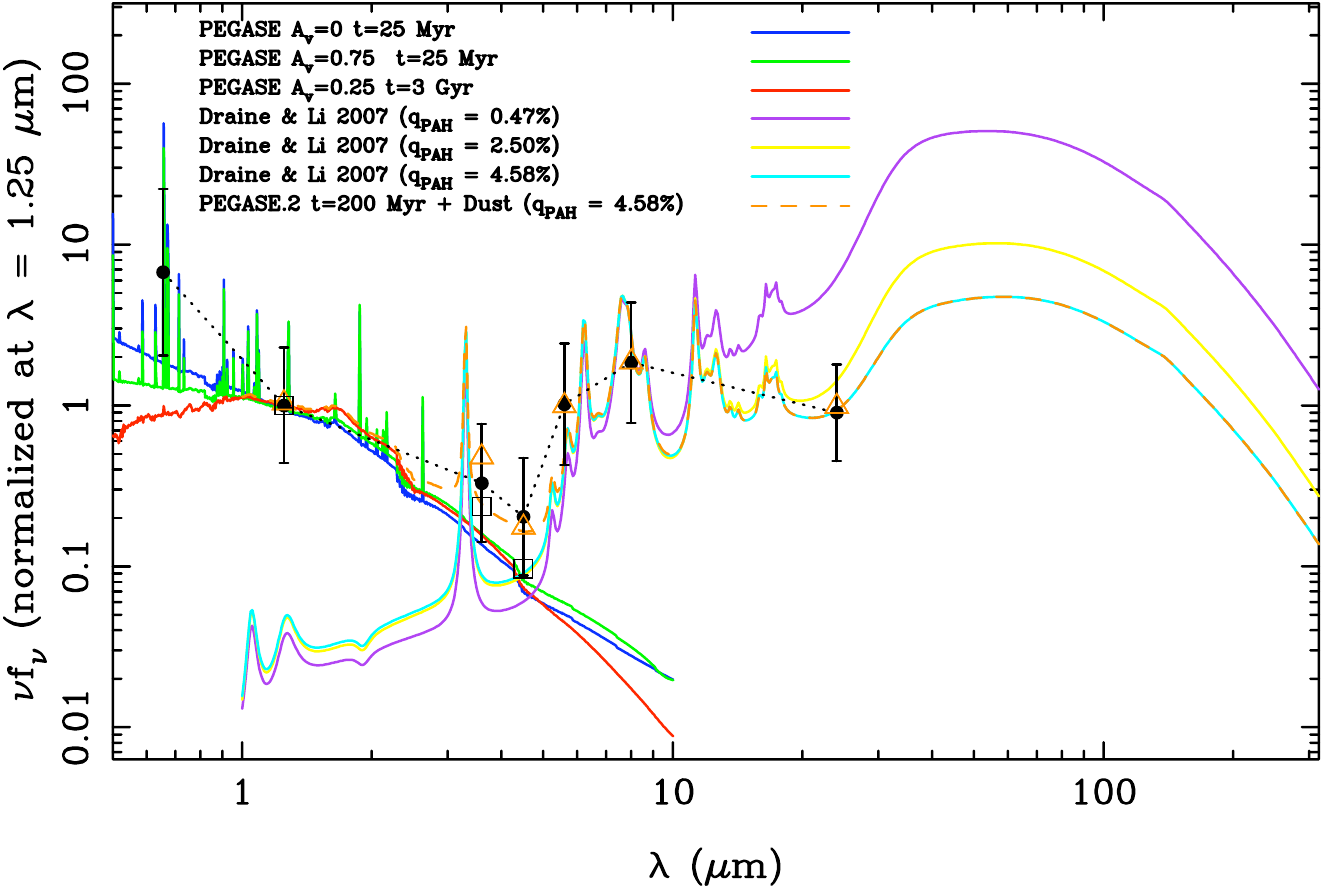}
\caption[The average 1-24\,\micron~spectral energy distribution of a star-forming environment]{Using the linear fits from Figure~\ref{fig:sbtosfr} (which scale with the star formation rate), a spectrum of broadband near/mid-infrared emission is plotted as solid black points (connected by the black dotted line), normalized at 1.25\,\micron. This average SED is a measure of the young stellar population and its associated dusty star-forming environment as seen in the near/mid-infrared as it is shown to scale with the intrinsic H$\alpha$ emission. The black open squares represent the colors (log$_{10}$ I$_{3.6}$/I$_{1.25} = -0.30\pm0.07$ and log$_{10}$ I$_{4.5}$/I$_{3.6} = -0.19\pm0.02$) of the majority of pixels in a galaxy (taken from the results of Figure~\ref{fig:pixels}). Shown for comparison are SEDs generated for stellar emission from PEGASE.2 models and dust emission models from \citet{dra07a}. The stellar models from PEGASE.2 are normalized at 1.25\,\micron, while the dust models, varying in PAH mass fraction ($q_\mathrm{PAH}$), from \citet{dra07a} have been normalized at the IRAC 8\,\micron~band. The sum of the young PEGASE model (blue) and $q_\mathrm{PAH}=4.58$\% (cyan) dust/PAH model is shown as the orange dashed curve. The open triangles represent the flux densities of this model in the 2MASS and \textit{Spitzer}/IRAC filters and are consistent within errors to the observed average SED data points (black solid dots).}
\label{fig:sed}
\end{center}
\end{figure*}

The flux ratios between H$\alpha$ and other wavelengths is obtained from the linear fits (and resulting  fitting errors) shown in Figure~\ref{fig:sbtosfr} and tabulated in Table~\ref{tab:coefs}. The SED points obtained in this way are then normalized to the $J$-band (at 1.25\,\micron) and plotted as black dots in Figure~\ref{fig:sed}. The black open squares represent the colors (log$_{10}$ I$_{3.6}$/I$_{1.25} = -0.30\pm0.07$ and log$_{10}$ I$_{4.5}$/I$_{3.6} = -0.19\pm0.02$) of the majority of pixels in the ensemble of galaxies. In Figure~\ref{fig:pixels}, this corresponds to the peak NIR color in the distribution panel (middle panel) and represents the NIR colors for non star-forming environments consisting of intermediate and evolved aged stellar populations. These NIR colors are consistent with the average typical NIR colors found for the integrated photometry of the SINGS sample \citep{dal07}, as well as for evolved galaxies in our previously studied high-z sample \citep{men09}. For reference, three SEDs from stellar population models using PEGASE.2 are plotted for a galaxy with an exponentially declining star formation history with an e-folding time of 500\,Myr at ages of 25\,Myr (blue and green lines) and 3\,Gyr (red line). The green spectrum is for a 25\,Myr old galaxy with dust extinction in the visible of A$_{v}=0.75$, while the blue spectrum has no dust extinction applied. These galaxies are dominated by the emission from the youngest stellar populations and is, in essence, a starburst stellar populations modeled according to stellar population synthesis. Contribution from non-stellar emission, which is known to be signficant for starburst galaxies at NIR and MIR wavelengths is not included. The older spectrum (red) is modeled with 0.25 mags of visual extinction. As emphasized earlier (in \S\ref{s:mot}), the stellar emission beyond 1\,\micron~is not dependent on the star formation history of the galaxy and near-infrared colors for a galaxy that do not have emission from PAH and/or dust molecules are independent of a galaxy's age. 

The average spectrum effectively measures the broad spectral energy distribution of a young stellar population and its associated dusty star-forming environment as the fits were made to minimize the contribution from emission from evolved stellar populations (except at H$\alpha$ where all stellar emission is effectively removed). It is evident that beyond $\sim2$\,\micron~the stellar SED does not provide sufficient emission to account for the observations. Although SPS models are extremely uncertain at these wavelengths, due to poor knowledge of stellar evolutionary phases and lack of good spectral templates, this conclusion is reinforced by the fact that typical NIR colors for stellar populations not contaminated by star-forming environments are way lower than those which are found in star-forming environments. Emission due to dust heated by the young stellar population is the obvious culprit. We plot for reference dust models, scaled to the 8.0\,\micron~flux density, from \citep{dra07a} for three PAH fractions of $q_\mathrm{PAH} = $0.47\%, 2.50\% and 4.58\% (these models assume $<U>_\mathrm{min}$ = 3.00 and $u_{\mathrm{max}}=1000$, see  \cite{dra07a} for details, and are shown for reference, not as best fits of the average SED). 

The orange dashed curve is the sum of the PEGASE.2 star-forming model (blue line) scaled to the 1.25\,\micron~flux density and the dust model with $q_\mathrm{PAH} =$ 4.58\% scaled to the 8.0\,\micron~flux density. The open triangles represent the flux densities of this composite model. The model flux densities are consistent with the average SED derived from the linear fits measured in this analysis. The scaling of the near/mid-infrared spectrum with the star formation rate suggests this region of the SED for a galaxy can be incorporated into galaxy SED models as a spectral component that scales linearly with the star formation rate.

\begin{deluxetable}{ccc}
\tablecaption{\small{Linear scaling of I$_{\lambda}$ with I$_{\mathrm{H}\alpha,int}$}}
\tablecolumns{3}
\tablewidth{0pc}
\tabletypesize{\small}
\tablehead{
	\colhead{$\lambda$ (\micron)} 		&
	\colhead{slope$_\mathrm{median}$}	 	&
	\colhead{slope $_\mathrm{range}$}	 	
}
\startdata 
 0.66 & 0.22 & 0.05 - 0.55 \\  
 1.25 & 0.06 & 0.02 - 0.14 \\  
 3.60 & 0.06 & 0.02 - 0.14 \\  
 4.50 & 0.05 & 0.02 - 0.10 \\  
 5.60 & 0.28 & 0.11 - 0.61 \\  
 8.00 & 0.73 & 0.29 - 1.46 \\  
24.00 & 1.07 & 0.44 - 1.72 \\  

\enddata
\label{tab:coefs}
\end{deluxetable}
%

%



\subsection{The NIR excess due to circumstellar (disk) emission}

\begin{figure}[tbp]
\begin{center}
\plotone{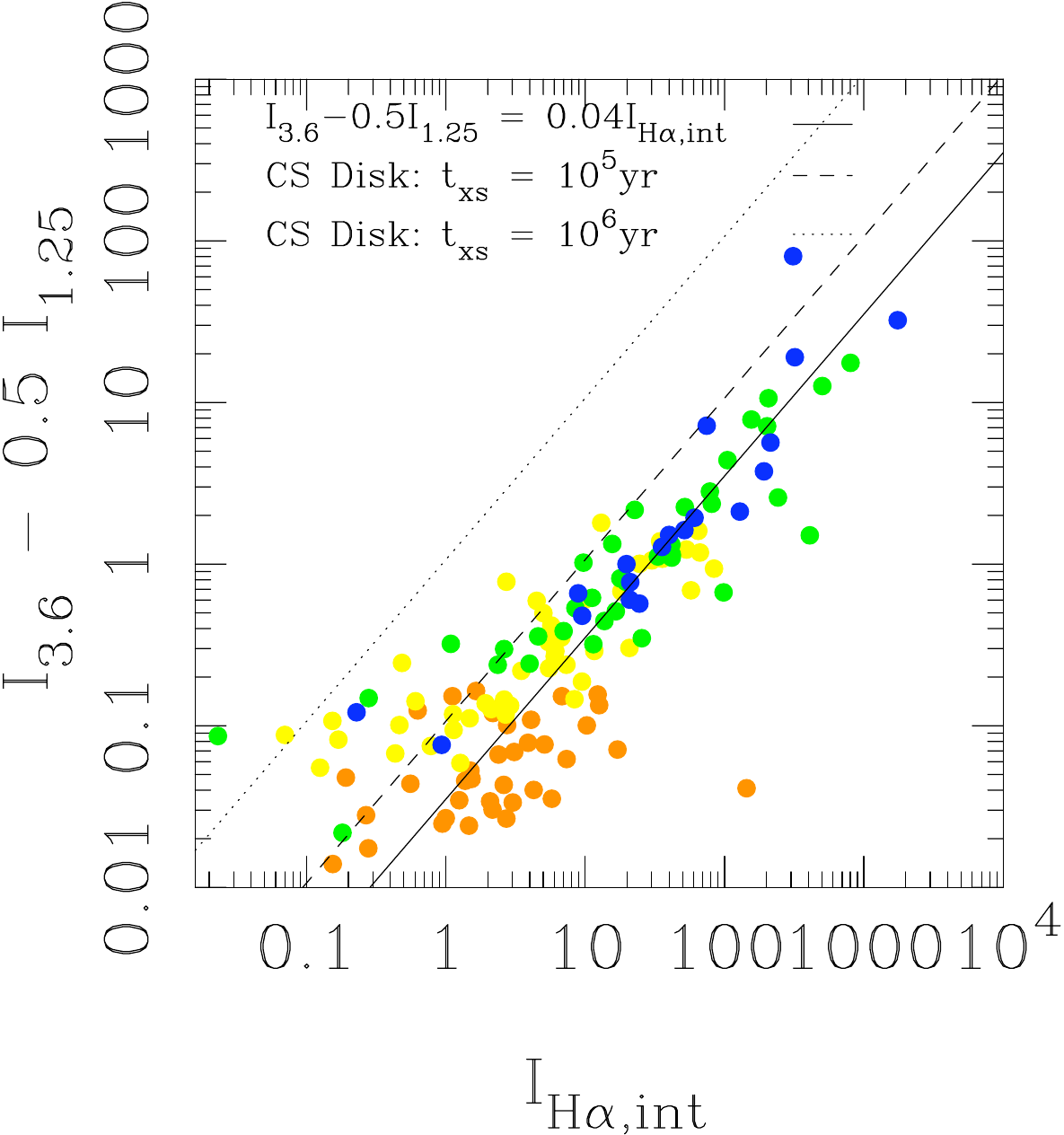}
\caption[The correlation between excess NIR emission and the star formation rate]{Excess NIR emission is plotted by subtracting the contribution of stellar emission at 3.6\,\micron~using emission at 1.25\,\micron. The excess emission is shown to correlate with the star formation rate with a median slope of 0.0435(+0.1092/-0.0311). The dashed and dotted lines is the equation for NIR excess emission due to circumstellar disk emission from \citet{men09}, as applied in Eq.~\ref{eq3:disk} and \ref{eq:sfr}, for disk lifetimes of $10^6$ and $10^5$ years respectively.}
\label{fig:xs_vs_sfr}
\end{center}
\end{figure}

A hypothesis for the origin of the NIR emission we have identified was proposed in \citet{men09}, namely that the emission is associated with circumstellar emission of dust around massive young stars. We can measure the excess NIR emission at 3.6\,\micron~by subtracting the contribution of stellar light from the bandpass. We use the ratio between emission in the near-IR of $\log$\,I$_{3.6}$/I$_{1.25} = -0.3$ (I$_{3.6}$/I$_{1.25}  = 0.5$) to measure the excess emission in the NIR and plot this as a function of the intrinsic H$\alpha$. A linear relation exists between the NIR excess (stellar subtracted) emission and is shown in Figure~\ref{fig:xs_vs_sfr}. A linear fit to the data is plotted as a solid line. The fit has a slope of 0.0435 (+0.1092/-0.0311).

We can use the quantitative model we developed in \citet{men09} to see if the excess emission at 3.6\,\micron~can be explained by the same model. A relationship between NIR emission from circumstellar disks from the model of \citet{dul01} was applied to the integrated light of a galaxy as:

\begin{equation}\label{eq3:disk}
L_{\mathrm{NIR}}(\rm{SFR})_{disk} = 350\,L_\odot / M_\odot \left( \frac{\mathrm{SFR}}{M_\odot/yr} \right) \left( \frac{t_{excess}}{yr} \right).
\end{equation}
\vspace{0.2cm}

\noindent this equation requires a SFR in $M_\sun$/yr and a value for the timescale of the circumstellar disk emission, $t_{excess}$.  The calculation assumed an Initial Mass Function from \citet{kro01} and as a result requires the same IMF assumption when converting the intrinsic H$\alpha$ intensity into a SFR. 

We use the standard conversion from \citet{Kennicutt:1998p693}, except an additional factor of 0.55 is applied to convert from an IMF from \citet{sal55} to an IMF from \citet{kro01}.

\begin{equation}\label{eq:sfr}
\rm{SFR} (M_\sun/yr) = 0.55 \times 7.942 \times 10^{-42}~{\rm erg\,s}^{-1}~L(H_{\alpha,int})
\end{equation}
\vspace{0.2cm}

The relation resulting from this model is plotted in Figure~\ref{fig:xs_vs_sfr} for two timescales of $10^5$ and $10^6$\,years, which cover the range of timescales in which circumstellar disks are found locally to survive. In general the model over predicts the emission, but is within an order of magnitude which is encouraging given the model's simplicity. This suggests that the excess NIR emission seen in nearby young massive stars \textit{can} explain the NIR colors seen in star forming regions in the SINGS galaxies and may be due to circumstellar dust heated to its sublimation temperature.

\subsection{AGN activity}

Do active galactic nuclei (AGN) cause red NIR colors? There is a trend for the NIR excess emission regions to be associated with regions of star formation. For many of the normal spiral galaxies in the sample, the regions of NIR excess spatially correlate with H$\alpha$ bright regions, found to be located along the spiral arms. Similar knots are seen in the irregular galaxies in the sample, but are found dispersed throughout these galaxies. In some galaxies  (both irregular and spiral types), some of these regions are found in the nuclear region of the galaxy. 

Table~\ref{tab:summary} contains a column which identifies whether a galaxy may harbor an active galactic nucleus using spectral indicators \citep{dal06}. We investigate each of the 24 galaxies with an AGN and find that in a third of the cases, the nuclear regions of galaxies do correspond to redder NIR colors. For 5 galaxies (NGC\,1266, NGC\,3031, NGC\,3198, NGC\,4736 and NGC\,7552), their nuclear regions have very red NIR colors (I$_{3.6}$/I$_{1.25} > 1$ ). The pixels in these regions have both high H$\alpha_{int}$ emission and high dust extinction. The red NIR colors may either be related to the AGN itself (or dust heated by the AGN); or could also be from optically thick dust extinction in starburst regions as the NIR colors in these nuclear regions are similar to NGC\,1366, an opaque starburst galaxy \citep{rou06}. 

\section{Conclusion}

The near-infrared, at 1-5\,\micron, marks a transition in the spectral energy distribution of galaxies. At $\sim$1-2\,\micron, emission from longer-lived low-mass stars from evolved populations remains bright, tracing the build up of mass in a galaxy. However, these wavelengths also trace currently star forming populations, which emit brightly in both the UV, optical and near-infrared. Near-infrared colors at 1-5\,\micron, if dominated by stellar emission, are predicted to be independent of the age of a stellar population. However, in star-forming environments, the contribution from Polycyclic Aromatic Hydrocarbons and dust grain emission becomes important beyond 2\,\micron, making the near- and mid-infrared a complicated but rich region of a galaxy's spectral energy distribution. 

In this paper, we have analyzed the pixel-by-pixel near-infrared colors for a sample of 68 galaxies from the \textit{Spitzer} Infrared Nearby Galaxy survey (SINGS; \citealt{ken03}) and Large Galaxy Atlas (LGA; \citealt{jar03}). Specifically we use images at 1.25, 3.6, 4.5, 5.6, 8.0 and 24\,\micron, as well as continuum subtracted H$\alpha$ images for 56 galaxies. Using the H$\alpha$ and 24\,\micron~images, we were able to create images representing the spatially resolved dust extinction (at $A(H\alpha)$) and the intrinsic H$\alpha$ intensity. The pixels in each galaxy were categorized into different regions based on a near-infrared color selection. The majority of pixels for each galaxy in the sample are found to have constant NIR colors, as is predicted for main-sequence stellar emission, while the reddest NIR pixels correspond to regions of star formation, which usually trace the galaxy's spiral arms structure. Our analysis suggests that very red NIR colors are produced by emission, not dust extinction.

For each NIR color cut in our galaxy sample, we measured the average intensities at 1.25, 3.6, 4.5, 5.6, 8.0 and 24\,\micron~for each galaxy. We compared these to the average dust extinction and star formation rate in each region and found a linear correlation between the emission at all bands from $\sim$1-8\,\micron~with the average H$\alpha_{int}$ intensity (and consequently the star formation rate). The strong correlation between near-infrared and PAH emission from 1-8\,\micron~and star formation suggests the average spectrum derived from this dataset is representative of a young stellar population and its associated dusty star-forming environment and can be incorporated to galaxy SED models very simply as a component directly scaled to the star formation rate of a stellar population.

\acknowledgements This analysis relied entirely on archival data. We would like to thank the SINGS collaboration \citep{ken03} and the Large Galaxy Atlas team \citep{jar03} for publicly sharing their datasets.

\bibliography{thesis}

\end{document}